\documentclass{article}
\usepackage[utf8]{inputenc}
\usepackage{graphicx}
\usepackage{multicol}
\usepackage{authblk}
\usepackage[margin=0.8in]{geometry}
\usepackage{array}
\usepackage{hyperref}
\usepackage[htt]{hyphenat}
\usepackage[sorting = none, backend = bibtex, style=numeric-comp]{biblatex}
\title{Cybersecurity Threat Hunting and Vulnerability Analysis Using a Neo4j Graph Database of Open Source Intelligence}

\author[1]{Elijah Pelofske\thanks{E-mail: elijah.pelofske@protonmail.com}}
\author[1]{Lorie M. Liebrock}
\author[2]{Vincent Urias}
\affil[1]{New Mexico Cybersecurity Center of Excellence, New Mexico Tech}
\affil[2]{Sandia National Laboratories}

\date{\vspace{-6ex}}

\begin{document}

\maketitle

\begin{abstract}
Open source intelligence is a powerful tool for cybersecurity analysts to gather information both for analysis of discovered vulnerabilities and for detecting novel cybersecurity threats and exploits. Here, we present a Neo4j graph database formed by shared connections (shared sub-string matches) between open source intelligence text including blogs, cybersecurity bulletins, news sites, antivirus scans, social media posts (such as Reddit and Twitter), and threat reports. These connections are comprised of possible indicators of compromise (IP addresses, domains, hashes, email addresses, phone numbers), information on known exploits and techniques (CVEs and MITRE ATT\&CK Technique IDs), and potential sources of information on cybersecurity exploits such as twitter usernames. The construction of the database of potential IOCs is detailed. Examples of utilizing the graph database for querying connections between known malicious IOCs and open source intelligence documents, including threat reports, are shown. We show that this type of relationship querying can allow for more effective use of open source intelligence for threat hunting, malware family clustering, and vulnerability analysis. We show four specific examples of interesting connections found in the graph database; the connections to a known exploited CVE, a known malicious IP address, a malware hash signature, and a portable executable shared resource file. 
\end{abstract}


\section{Introduction}
\label{section:introduction}

Open source intelligence offers an extraordinary amount of information that a cybersecurity analyst can use for threat detection, mitigation, and analysis \cite{glassman2012intelligence, evangelista2021systematic, steele2007open, piplai2020knowledge, 9458871, 10.1007/978-3-030-59621-7_2}. However, open source intelligence contains a large amount of noise (i.e., irrelevant information) and most importantly the scale of the data is too large to be useful in its raw form. To this end, automating the process of finding indicators of compromise and relevant relationships between the indicators of compromise, has become increasingly important \cite{catakoglu2016automatic, kazato2020improving, 9458871}. The central idea utilized in this study is forming a network of associations between open source intelligence documents and \emph{potential indicators of compromise} (IOCs) that exist in the open source intelligence text. The term \emph{potential IOC} is important because it specifies that the text is potentially relevant (for example, discussing usage of a new piece of malware) and can be unstructured natural language text, but there is a pattern match in the data that does fit a particular form (for example, an IP address, or a common vulnerabilities and exposures (CVE)). However, given the nature of open source text and information, there exist many \emph{false positives} - e.g., correlations that exist in open text that do not actually have a semantic reason for happening or are not important for the type of data we wish to extract (in this case, cybersecurity relevant text). Note that these types of correlations are not false positives in the traditional sense of predictor based models. Instead, these types of correlations are real, in the sense that the text is the same, but \emph{semantically} the text strings may be referring to two different concepts. 

Graphs are a natural way to express these types of higher order connections and are used in a variety of cybersecurity contexts \cite{10.1145/3460120.3485353, 10.1145/3469379.3469386}. These types of networks, when they are intended for providing semantic meaning between heterogeneous data types are also referred to as \emph{knowledge graphs} \cite{10.1145/3442520.3442535, 10.1007/978-3-030-92836-0_22, https://doi.org/10.13140/rg.2.2.27340.95367, https://doi.org/10.13140/rg.2.2.27340.95367, liu2022reviewknowledgegraphapplication}. In order to create a graph representation of a large amount of open source intelligence, we utilize Neo4j which provides a visual interface to search the graph, allows a number of users to interact with the data, and also provides an efficient query time for the database to interact with other analysis systems or to simply query and display the raw open source intelligence (OSint) text documents that are connected to a relevant exploit or IOC. Neo4j graph databases have been used in other domains for the purpose of storing and querying data structures with complex networks \cite{miller2013graph, guia2017graph, pokorny2015graph, huang2013research} including social network analysis \cite{warchal2012using} and typhoon disaster knowledge \cite{liu2020construction}. We utilize Neo4j because it is a reasonable choice for an existing graph database implementation -- in particular, it is efficient, open source, and there exist Python 3 libraries for interacting with the database. Utilizing graphs in order to better evaluate relevant connections that exist in a large dataset is a subject of considerable interest~\cite{noel2016cygraph, jia2018practical, joslyn2013massive}.

\begin{figure*}[h!]
    \centering
    \includegraphics[width=0.9\textwidth]{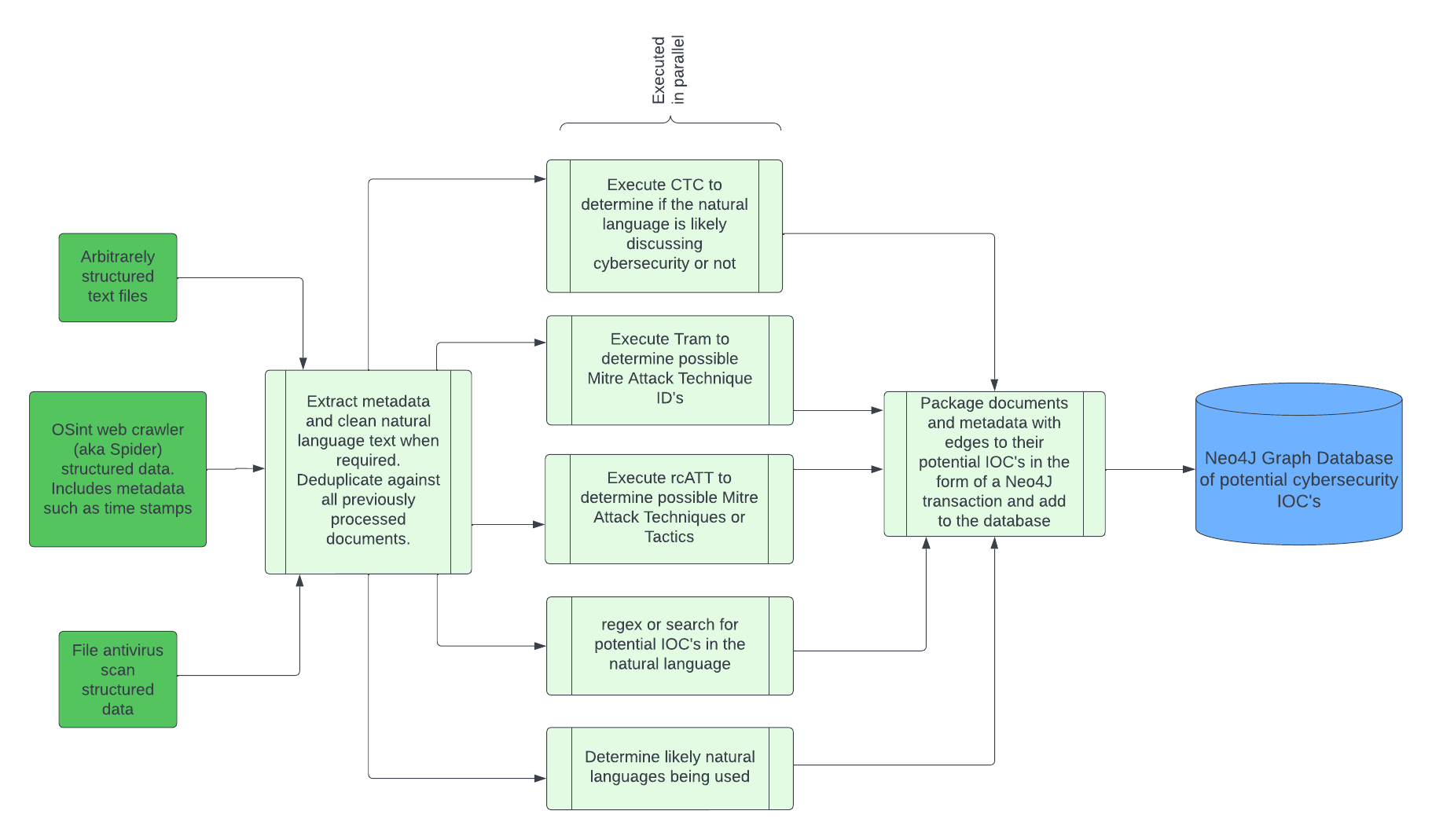}
    \caption{IOC Neo4j graph database construction workflow diagram. }
    \label{fig:workflow_diagram}
\end{figure*}

In this study we outline a methodology which consists of collecting and aggregating open source intelligence, in conjunction with antivirus scan results and presenting the information contained in this data in the form of a \emph{network} or a graph. Specifically, edges represent a connection between a potential IOC and a document. The challenge is that most open source intelligence text from the web is unstructured therefore analyzing the text for relevant pattern matches is a means to extract the potential IOCs. Here a document is simply a collection of natural language or data - for example a single tweet from Twitter could be a \emph{document}. The knowledge graph that is formed by the documents, potential IOCs, and vulnerability IDs form a network with multiple types of relations (edges) -- making the constructed graph database a specific type of multidimensional network. Furthermore, metadata is included with the node document when constructing the database - several machine learning (ML) algorithms identify whether the text is discussing cybersecurity and specifically what type of exploit techniques are being discussed. This type of metadata is important because it can serve as a signal for the quality of the given node document, and can be a fast way to summarize the high-level information about a document (such as the natural language present in the document) when searching through a large graph database halo of nodes. A piece of open source intelligence text gathered from social media will have a higher confidence of being relevant if the machine learning algorithm tells us it is likely discussing a cybersecurity topic. In Section~\ref{section:methods} we detail the construction of this type of cybersecurity intelligence graph, instantiated as a Neo4j graph database. 

The underlying idea of this study is to create a graph database structure where edges between two nodes are formed by shared string matches within two open source documents, and each node is a single open source document (which could be from a variety of sources, such as a threat report or a web page). Our primary contributions are describing an open set of methods and techniques to create this type of indicator graph database. We describe and give examples of real world data found in very noisy internet natural text sources, such as 4chan and Reddit, which makes the reliability of the ingested information hard to assess automatically. We also report on aggregate level graph statistics on CVE nodes within the graph database, compared to their base severity scores.

In Section~\ref{section:results} we give specific examples of where this type of open source intelligence graph database can provide useful information and facilitate analysis in cybersecurity. Specifically, we detail examples where indicators of compromise can be found within the graph database, and we show where connections to other open source documents can give more information as to how that indicator is important. In Section \ref{section:results_malware_PE_shared_resource} we show how a shared PE resource file can be used to link multiple malware samples (likely due to the same developer creating these pieces of malware) that otherwise would not be connected by standard malware analysis tools. We conclude in Section \ref{section:discussion} with a discussion of what this type of approach to open source intelligence analysis can be used for and future research directions.

\subsection{Related Work}
\label{section:intro_related_work}

Refs.~\cite{mti3030062, Caballero_2023, takko2022knowledgeminingunstructuredinformation} have studied similar systems that use automated text mining, including unstructured text, to extract indicators of compromise or cybersecurity relevant information. Our study differs from these studies in several ways. The first is the parsing of entirely unstructured text, largely from various internet sources such as social media and blogs, to be ingested and the second is that this very general parsing is applied to a wide variety of data sources. The third is the variety of extracted datatypes used to create the graph database - this not only includes standard indicators of compromise, but also datatypes with more general context for known vulnerabilities and tactics (namely, CVE numbers and MITRE ATT\&CK ids). 

Our approach is based on creating a graph database, which is a type of knowledge graph, based on a collection of cybersecurity indicators. The approach of creating cybersecurity knowledge graphs has been extensively studied and appears to be a very successful means of organizing data as complex as threat intelligence~\cite{li2023k, 10.1007/978-3-030-92836-0_22, 10.1145/3442520.3442535, li2022attackgconstructingtechniqueknowledge, fieblinger2024actionablecyberthreatintelligence, li2022novel, sarhan2021open, husak2023provision}. Our approach is similar to prior knowledge graph studies, namely in that the key unit of information for cybersecurity analysis is indicators of compromise. However, we use a very broad set of indicators including several different types of direct string pattern matches.

Ref.~\cite{6805774} described a way to track the expansion of indicators in a graph, which is very directly what the present study demonstrates using a graph database. 

Our contribution is a complete description of the construction of a Neo4j graph database of cybersecurity indicators. The core mechanism that the graph database enables is searching based on potential IOCs, which then finds documents that contain those IOCs, and from there a local halo of associated documents and potential indicators and adjacent data can be explored. We show several clear examples from open source text of known IOCs having relationships within the graph database. Moreover, using CVEs as a good reference test case, we highlight several graph metrics such as the PageRank score of CVE nodes in the database, as well as showing an interesting lack of a linear correlation between CVE CVSS score and CVE node popularity. The CVE graph measures are statistically significant, spanning over 170,000 distinct CVEs, making those findings of independent scientific interest. Our graph database does not rival the scale or the infrastructure that commercial threat intelligence tools have, instead what we demonstrate is a specific approach to efficient open source intelligence parsing (using computationally efficient techniques such as direct string matching). We show that a moderate-scale open source text parsing system can be created, and this study describes the technical details and illustrative examples of the database containing real-world interesting connections.

\section{Methods}
\label{section:methods}

In this section we outline the automated pipeline used to construct the Neo4j database. In Section \ref{section:methods_data_sources} the open source intelligence data sources are defined, in Section \ref{section:methods_neo4j_database_construction} the database construction is outlined. 

\subsection{Open source intelligence data sources}
\label{section:methods_data_sources}

The open source intelligence documents that are fed into the system broadly fall into three categories; the three green starting blocks in Figure \ref{fig:workflow_diagram} show these categories. The first type of data is entirely unstructured text - typically this type of data is simply a threat report or a list of indicators or vulnerabilities. This option allows any collection of text sourced from any internet discussion to be analyzed for potential IOCs and to be added into the Neo4j database. Figure~\ref{fig:workflow_diagram} was created in Lucidchart. \footnote{\url{www.lucidchart.com}}

The second type of data comes from a system of web crawlers (the details of this system are outlined in ref.~\cite{9717379}). The text found by the web crawlers is typically similarly unstructured and can originate from a variety of sources such as Reddit, Twitter, blogs, cybersecurity bulletins, and news sites. However, the web crawler data also contains metadata including links that the crawler followed to this current site, time and date information, and cybersecurity keywords that were found when parsing the site. Therefore, this data is slightly more structured and is therefore parsed differently from arbitrary text; specifically the metadata is kept separate from the text that will be analyzed for potential IOCs, which is eventually added into the node information in Neo4j. 

The third type of data is structured antivirus scans of files that are potentially malware. This data is entirely machine readable and does not contain natural language information. Therefore, this data is also parsed differently to the other two data sources; these antivirus scans include hashes of files and the names of those files. Therefore, those specific fields (hashes and file names) that are present in the structured data are parsed for creating edges in Neo4j, but no other natural language analysis or pattern matching for potential IOCs is performed.

\subsection{Neo4j Graph Database Construction}
\label{section:methods_neo4j_database_construction}

\begin{table*}[h]
\centering
\begin{tabular}{ |c||c|c| }
 \hline
 Node type & Count of unique nodes & Count of edges \\ 
 \hline
 \hline
 Document & 2,128,992 &  \\ 
 \hline
 \hline
 MD5 Hash & 394,826 & 944,671 \\ 
 \hline
 SHA1 Hash & 323,321 & 990,817 \\ 
 \hline
 SHA256 Hash & 642,535 & 1,941,248 \\ 
 \hline
 SHA512 Hash & 18,339 & 28,112 \\ 
 \hline
 Malware name & 365 & 117,528 \\ 
 \hline
 APT name & 457 & 165,741 \\
 \hline
 Email & 85,396 & 953,757 \\ 
 \hline
 CVE ID & 174,668 & 1,313,206 \\ 
 \hline
 Twitter username & 174,143 & 402,899 \\
 \hline
 Phone number & 23,756 & 144,397 \\
 \hline
 IP address & 119,699 & 705,386 \\
 \hline
 domain & 214,720 & 3,116,508 \\
 \hline
 File name & 351,507 & 1,326,480 \\
 \hline
 MITRE ATT\&CK Technique ID & 445 & 21,018 \\
 \hline
 \noalign{\vskip 1mm}
\end{tabular}
\caption{Neo4j database statistics. The \emph{Count of edges} column in the table is the number of edges between the listed node type and Document nodes. For each non-document node type, the edges between document nodes and the unique nodes for that indicator are always labelled the same as the indicator; for example CVE nodes are connected to document nodes by edge labelled as \emph{CVE}. Those edge types are the counts displayed in the third column. Therefore, there are no edges between document nodes which is why the edge count in that cell is empty. }
\label{table:Neo4j_stats}
\end{table*}

Figure~\ref{fig:workflow_diagram} details the high level workflow that constructs the Neo4j database of potential IOCs. This entire system can operate continuously by reading in new data from the independent web crawler infrastructure and then adding relevant information into the Neo4j database. First, the data is parsed depending on its source, as outlined in Section \ref{section:methods_data_sources}. Each discrete input of text is referred to as a \emph{document} because this collection of text is all originating from the same source and is therefore logically linked, which could be helpful for identifying relationships when searching the database. As an example, a single document could be a piece of text from social media, such as a Reddit post or a tweet from Twitter. Any metadata that is associated with the input text is parsed at this point, to be added into the Neo4j database if a node representing this document is created. Importantly, once the raw text input has been extracted (this includes the JSON structured antivirus file scan data), a SHA256 checksum of the data is computed. This checksum is then checked against the checksums of all of the documents that are in the Neo4j database; and if the document is a duplicate, it is not added to the database. Next, several processes begin executing in parallel, all of which are designed to extract useful meaning from the natural language input. To this end, the antivirus file scan data is not parsed using the machine learning algorithms or language detection; however the potential IOCs such as filenames and hashes are extracted and used to create edge relationships in the database.

\subsubsection{Neo4j Graph Database Construction: Automated Machine Learning Models}

The Cybersecurity Topic Classification (CTC) tool, which is comprised of multiple machine learning algorithms trained to detect cybersecurity vs non-cybersecurity discussions from social media and developer forum English text sources~\cite{pelofske2021enhanced}, is executed on the text. The output of this algorithm is simply three states - either it is likely cybersecurity related text, or not, or there was not enough data (e.g., English words) to make a decision. 

The Reports Classification by Adversarial Tactics and Techniques (rcATT) machine learning python tool \cite{legoy2019retrieving} is also executed on the text. This tool gives a list of likely MITRE ATT\&CK tactics and techniques \footnote{\url{https://attack.mitre.org/}} that were mentioned in the text. The MITRE ATT\&CK framework provides a consistent basis for tracking cybersecurity techniques \cite{kwon2020cyber, att2021mitre, al2020learning, kuppa2021linking, https://doi.org/10.48550/arxiv.2211.06495, https://doi.org/10.48550/arxiv.2211.06500}. The rcATT tool was trained on threat report text and therefore the error rates on non-threat report documents is expected to be high. The tool is applied uniformly to all documents because it is not necessarily known a-priori what the exact semantic content of the document is (e.g., whether it is a threat report, or a news report, or entirely non-cybersecurity related). It is generally expected that based on the training data used to create the machine learning models in the tool, cybersecurity content (e.g., documents where CTC returned True) will have higher accuracy results in regards to detecting discussions of specific MITRE ATT\&CK tactics and techniques. 

The Threat Report ATT\&CK Mapper (TRAM) machine learning python tool\footnote{\url{https://github.com/center-for-threat-informed-defense/tram/}} is also executed on the document. TRAM, similar to rcATT, returns likely MITRE ATT\&CK techniques that were mentioned in threat reports. Therefore similarly to rcATT, this data will likely have high error rates for natural language text that is quite different from threat reports, but could be more accurate for cybersecurity related text. 

For the data coming from the web crawlers, the natural language text could be non-English. Having some signal to indicate when this occurs in the Neo4j database could be useful (for example if one just wants to query documents that are only English or Spanish text). The other reason that this signal is important is because all of the natural language machine learning algorithms vectorize the input text using a very broadly defined English dictionary - meaning that other languages are not used in these models, which means that their results will be very inaccurate and should not be used. Therefore, the python tool \emph{langdetect} is also executed on the text and the resulting language detection information is included in the node metadata. The machine learning models were applied to every document regardless of the language mixture, which means that ML outputs for the non-English documents in practice should not be considered. Future updates to this system should include more specific handling for multiple different natural languages, including ML parsing and vectorization within each language.

\subsubsection{Neo4j Graph Database Construction: Text Pattern Matches}

Lastly, pattern matches for all potential IOCs are performed on the text. With the exception of the structured antivirus file scans, where the potential IOCs that can be extracted can be done automatically, all of the other input text can be entirely unstructured. Therefore, simple pattern matching procedures are performed in order to find potential IOCs. Hashes (md5, sha1, sha256, sha512) are found by searching for high entropy hexadecimal text that fits within the required character length. File names are found by matching tokenized words (NLTK \cite{bird2009natural} was used for most of the tokenization procedures) that have a file ending that matches some standard file ending (for example \textit{.py} for python). Advanced Persistent Threat (APT) group names and malware names are all simply pattern matched against tokenized words in the text. Phone numbers, Email addresses, IP addresses, Twitter usernames, and domain names are all found by pattern match searching for known standard formats. Some simple checks are used to rule out pattern matches which do not fit the expected format of the data type and in the case of domain names, the top 1 million (Alexa top 1 million list) most searched domains are removed in order to reduce noise in the graph. Common Vulnerabilities and Exposures (CVEs) \cite{9874201} and the unique ID numbers of MITRE ATT\&CK techniques \cite{strom2018mitre} are also pattern matched for; both of these data types also follow a standard format which can be identified. Each of these pattern matches will correspond to an edge (e.g., a connection) between the document node that contained this data and a node representing that unique pattern match. This unique pattern match we broadly call a \emph{potential IOC}, but it can also simply be a unique ID to track a known vulnerability (for example a CVE), or it could be a potentially useful piece of information for connecting two document nodes but not be a malicious IOC.

\subsubsection{Neo4j Graph Database Construction Details}

Once all of this pattern matching and data processing has been completed, a number of \emph{Neo4j transactions} are applied in order to add the new data into the database. All of the processed documents have unique nodes created which contain several different components. The most important part of the node is the original raw text that was in the node document. Next, web crawler metadata is in the nodes which originated from web crawling - here the metadata includes the link, parent link, time stamps, keywords, and potentially other data such as checksum of the raw text. Next, some document nodes are antivirus scans - in these cases there is a large amount of metadata about the scan that was performed, but typically the data is not natural language. All nodes also have language detection metadata, if there was natural language text that could be processed. The natural language detection is useful for filtering nodes which contain only a specific language of interest, however as expected most of the document nodes are English text. The pattern matches that were found in the text that form the edges in the graph are also included as a segment in the node data. Lastly, the machine learning results from CTC, rcATT, and TRAM are also included as a separate data segment (if there is applicable English text that could be fed to these ML algorithms). Each of these components of node data can be used to filter for specific nodes which have specific attributes that are relevant for a specific task. Any documents which have no pattern matches are not added to the database as they would simply be degree 0 nodes. Next, for all unique pattern matches found in this set of documents, check if the nodes already exist in the database. If they do exist, do nothing, but if they do not exist, then create them. This is to avoid creating duplicate potential IOC or pattern match nodes. Lastly, edges are formed between the document nodes and the nodes representing the pattern matches found in those documents. Table~\ref{table:Neo4j_stats} shows the counts of nodes and edges in the database at the time of writing. Therefore, the constructed graph database is always bipartite where one partition is the node documents (i.e., the sources of information) and the other partition is the set of nodes representing different potential IOCs or vulnerability IDs that exist in the node documents. This structure allows users, or algorithms, to query for relationships based on potential IOCs (e.g., a path connecting two potential IOCs or the network of neighbors associated with a potential IOC) and then examine the sources of the information within the node documents, including machine learning and language metadata. 

Note that other types of cybersecurity relevant strings, besides those used in this study, could be searched for and created as a type of edge in such a graph database. The underlying idea of constructing a graph database based on extracted strings that match certain patterns can certainly be generalized to other datatypes. This set of string matches that we searched for in this study serve as a representative set of (known) cybersecurity relevant indicators and datatypes.

The computer platform on which the Neo4j database was created in our implementation required sufficient storage for all of the source documents (which is on the order of hundreds of gigabytes), at least 32 gigabytes of RAM to operate the database and handle building the database, and sufficient processor cores to perform multiprocessing when required (this is not a highly parallel processing intensive task and therefore on the order of 8 cores in total is sufficient). Of course, scaling this graph database text mining system to significantly larger datasets would require more compute resources. The data used to construct the graph database was collected from the years 2018 through 2022 at a university research program at New Mexico Tech. Although the graph database includes threat reports and natural language information from prior to these dates, most of the active scraping of information occurred in 2020-2021, and therefore we expect a majority of the correlations and news content that was gathered to be from these years.

The use of the hash checksums is motivated because these are standard malware signatures.  Advanced persistent threat (APT) names and malware names are used commonly in threat reports and discussions of threat group activity - and both malware and APT groups can have multiple different names associated with them, which makes the addition of these known names into the database potentially useful. Email addresses, Twitter usernames, phone numbers, IP addresses, and domain names are all standard indicators of compromise. File name can be a useful indicator of compromise if, for example, a malware file name is unique, but otherwise can result in a very high degree node with many connections due to a commonly used file name. CVE IDs and MITRE ATT\&CK Technique IDs are both useful for extracting information from threat reports and vulnerability reports for post incident analysis, which can then potentially be connected to other interesting regions of the graph.

\section{IOC Connections in the Graph Database}
\label{section:results}

This section shows several specific visual examples of the structure of the database indicates what is available to the user when searching for IOCs or vulnerability IDs. The Neo4j database construction is entirely automated, but here we highlight the use of manual Neo4j graph exploration to visually illustrate to the reader concrete examples of known IOCs, and how they are related in the database. The hard part of cybersecurity analysis is aggregating information, in this case open source unstructured text, into a meaningful form that can provide a network of related documents and potential indicators of compromise. Therefore, this set of examples that we show is motivated by giving a demonstration of a walkthrough of gradually expanding out from a node to find notable connections in the graph. In particular, because of the nature of this graph database creating potentially many false positive correlations (such as edges which are not cybersecurity relevant), we illustrate some specific examples that emulate how a human user would interact with the database. Concretely, there are specific types of indicators which are significantly more reliable data types - namely data which is unique such as hash checksums, twitter usernames, CVE numbers, and MITRE ATT\&CK Technique IDs. The remaining types of extracted data from the text mining can have much higher false positive edges created in the database because of the nature of those strings (for example, IP addresses and phone numbers being sequences of numbers means that un-structured parsed text from the internet is more likely to contain an incorrect match to one of those datatypes). Note that there are \emph{many} documents in the graph database, and many do not have interesting content; these selected examples have clear relevant content.

In Section \ref{section:results_visual_md5_hash} we show the graph connectivity around an md5 hash of a known malware file from the WannaCry ransomware. In Section \ref{section:results_visual_IP_address} we show the graph connectivity surrounding an IP address known to be a command and control server for the Qakbot trojan. In Section \ref{section:results_visual_CVE_exploited} we show the graph connectivity around a CVE that is known to be exploited in the wild. 

Section \ref{section:results_malware_PE_shared_resource} shows a specific case where a sha256 hash of a contained resource was in several antivirus scans of different portable executable malware. This shared contained resource is reasonably unique, and indicates software re-use during the development of otherwise seemingly unconnected malware samples. 

In Section \ref{section:results_CVE_degrees} we show the distribution of CVE node degrees in the Neo4j database compared to their Common Vulnerability Scoring System (CVSS) scores, and show that for reputable open source cybersecurity sources (e.g., threat reports), there is a weak-to-median linear relationship between the two. Lastly, in Section \ref{section:CVE_page_rank} the Neo4j Graph Data Science implementation of PageRank is applied to the graph database, allowing a ranking of the most influential CVEs across the entire database. 

\begin{figure}[h!]
    \centering
    \includegraphics[width=0.42\textwidth]{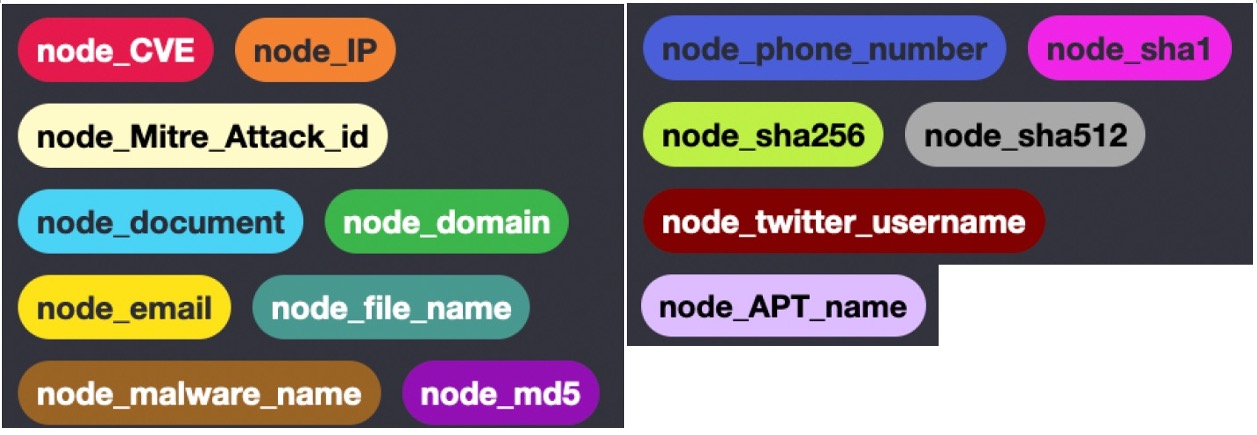}
    \caption{Node and edge coloring legend for potential IOCs. }
    \label{fig:node_legend}
\end{figure}

For the visual graph examples we query the database with the \emph{Neo4j browser} using the Cypher query language\footnote{\url{https://neo4j.com/developer/cypher/}}. The node and edge coloring's encode the following information. Document nodes are large cyan nodes, SHA1 nodes are magenta, SHA256 nodes are lime, SHA512 nodes are grey, MD5 nodes are purple, file name nodes are teal, email address nodes are yellow, IP address nodes are orange, malware name nodes are brown, Twitter usernames are maroon, APT name nodes are lavender, domain name nodes are green, phone number nodes are blue, CVE number nodes are red, and MITRE ATT\&CK technique IDs are beige. The edge coloring matches the node coloring; for example an edge connecting a CVE to a document node where it was mentioned will also be colored red. Figure \ref{fig:node_legend} shows the node and edge coloring legend. In order to reduce visual clutter in the graph connection structure plots, if there are near duplicate documents from the web crawlers which have the same connections to a group of nodes, we manually remove all but one of the duplicate document nodes. Here, near duplicate means that the two documents (meaning, a string of text) did not have exactly the same hash checksum, which is typically because of small text changes to the website text over time, however the vast majority of the text is still repeated. Query times for these examples require on the order of seconds of wall clock time to return the query.

\begin{figure}[h!]
    \centering
    \includegraphics[width=0.2\textwidth]{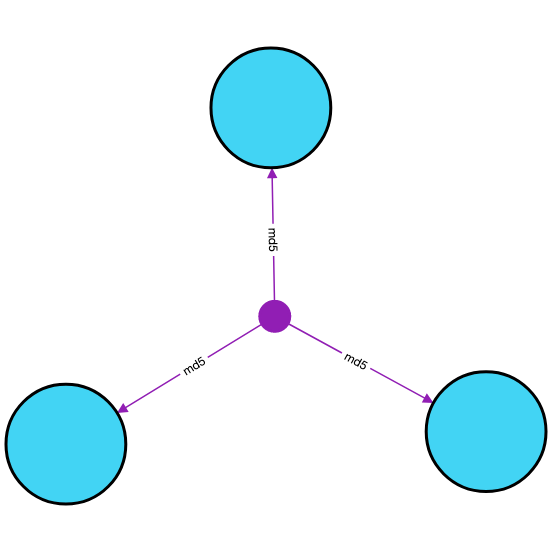}
    \caption{Degree 1 connections associated with the md5 hash \textbf{84c82835a5d21bbcf75a61706d8ab549} from the Neo4j graph database. This simple graph structure shows that within the current database this hash checksum is mentioned in exactly 3 open source documents, shown as light blue nodes. }
    \label{fig:md5_connections_1}
\end{figure}

\begin{figure*}[h!]
    \centering
    \includegraphics[width=0.6\textwidth]{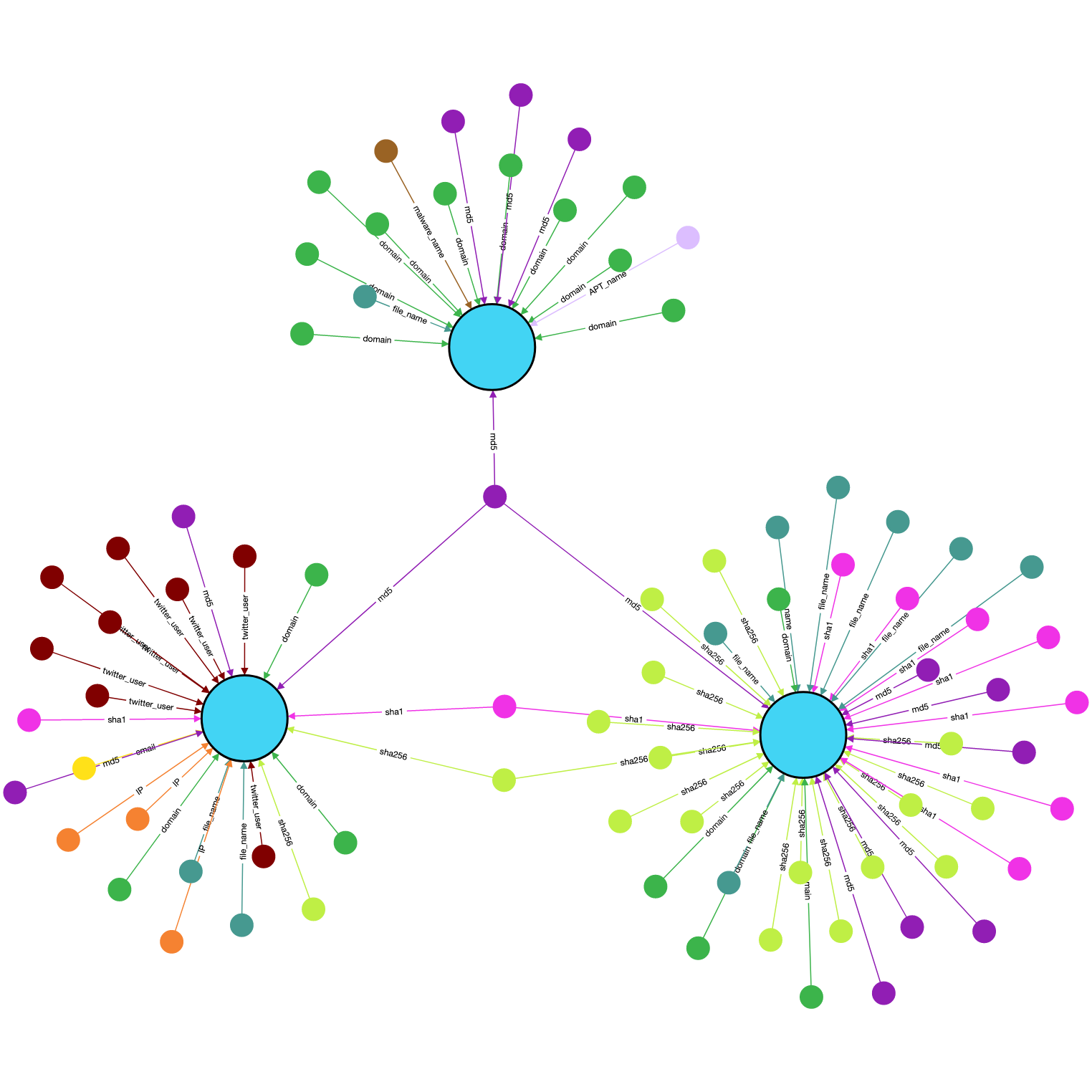}
    \caption{Expanded connections from Figure \ref{fig:md5_connections_1}; degree 2 connections out from the md5 hash \textbf{84c82835a5d21bbcf75a61706d8ab549}. This step now shows that each of the 3 open source documents that contain this hash have a variety of extracted datatypes, including a SHA-1 hash node and a SHA-256 hash node that are shared between two of the open source documents. }
    \label{fig:md5_connections_2}
\end{figure*}

\begin{figure*}[h!]
    \centering
    \includegraphics[width=0.6\textwidth]{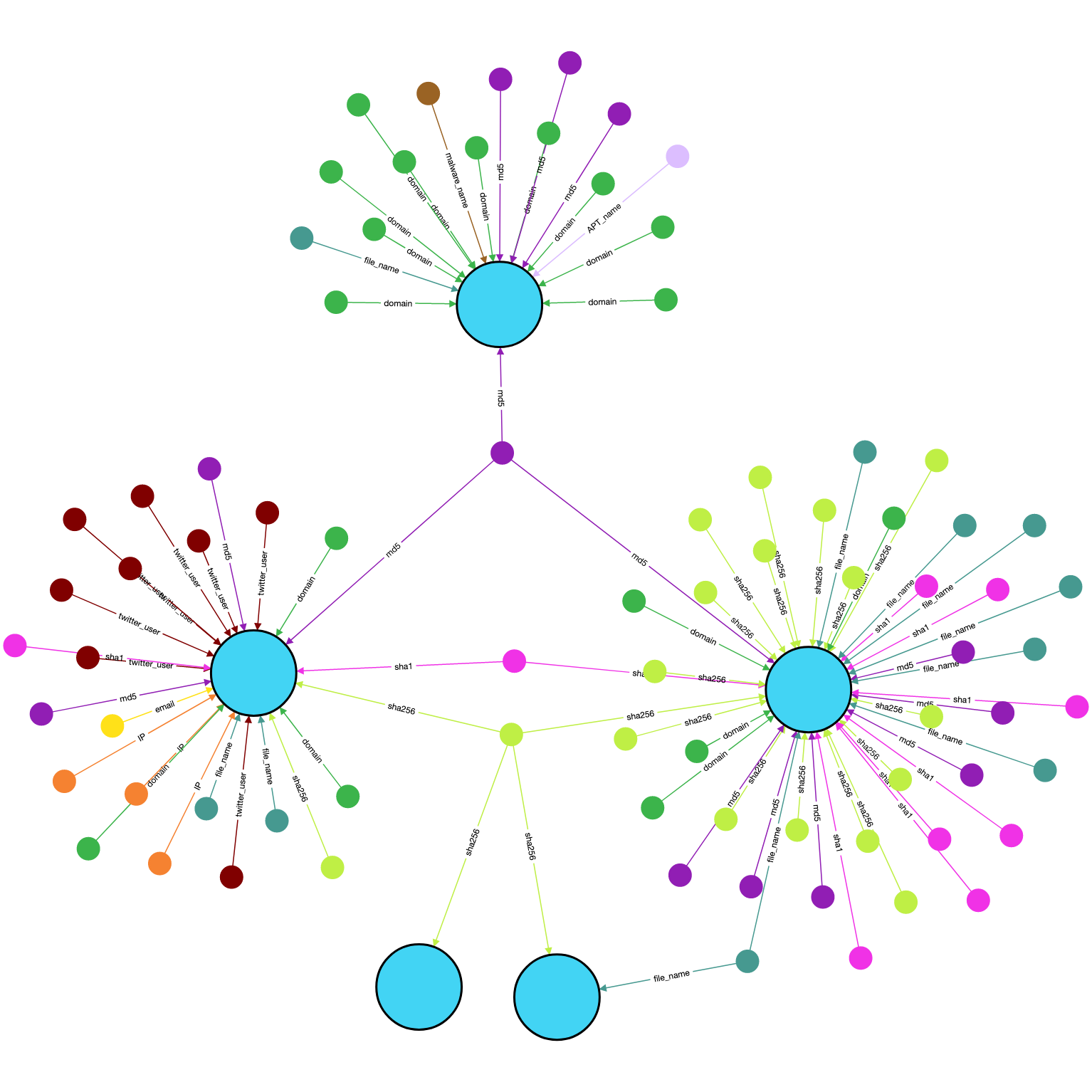}
    \caption{Expanding the neighbors for two of the hash node connections from Figure \ref{fig:md5_connections_2}, which were the only two shared nodes among the expanded neighborhood of the three original matched open source documents. Only one of these two shared hash datatype nodes had any further connections -- which turned out to be two other open source documents.  }
    \label{fig:md5_connections_3}
\end{figure*}

\begin{figure*}[h!]
    \centering
    \includegraphics[width=0.6\textwidth]{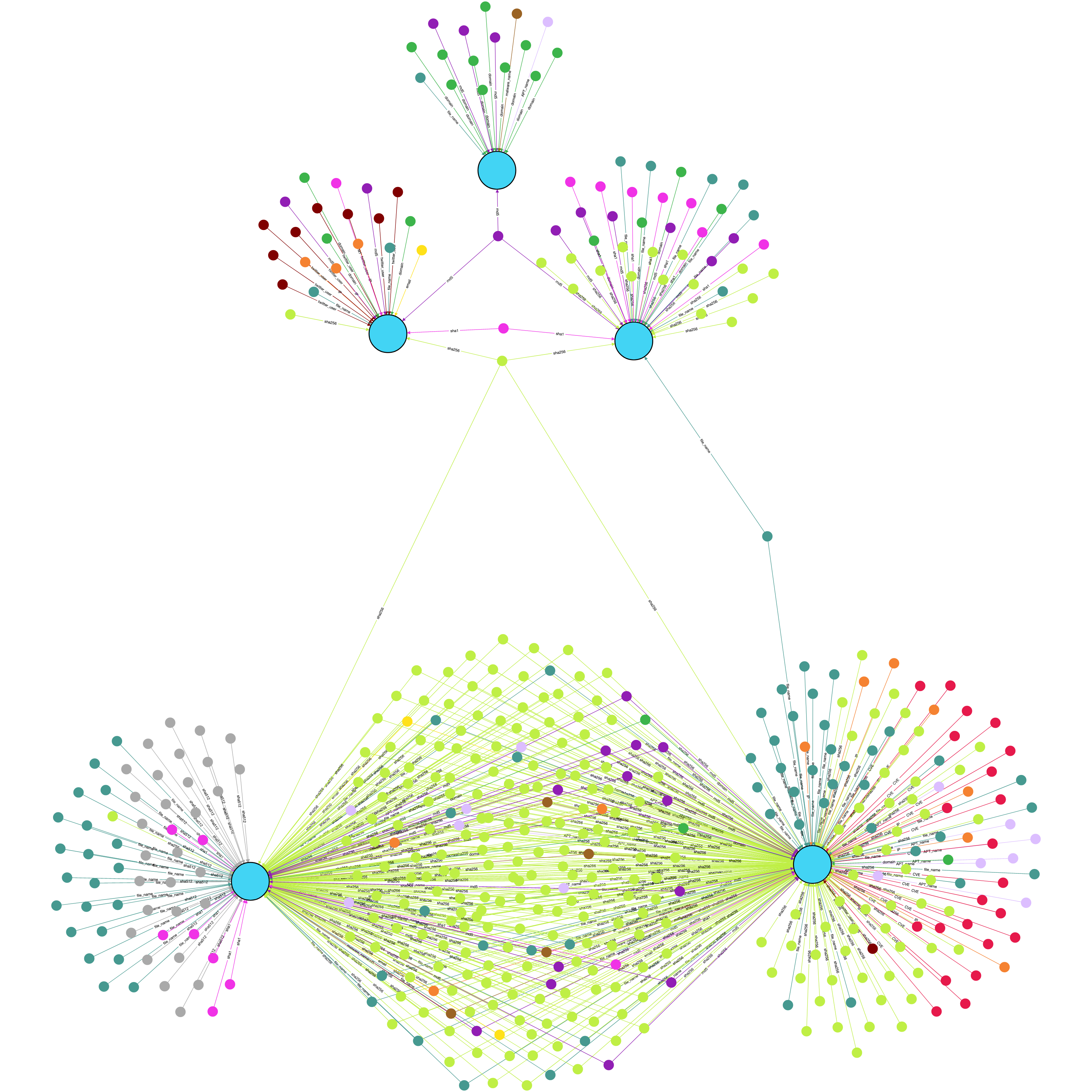}
    \caption{For this figure, begin by expanding the two threat report node connections that were connected to the two shared hash nodes in Figure \ref{fig:md5_connections_3}. Expanding the neighborhood of these two open source document nodes, which are threat reports, then revealed a large halo of extracted datatypes and indicators that were mentioned in these two threat reports. Using this chain of graph edge connections shown through Figures \ref{fig:md5_connections_1}, \ref{fig:md5_connections_2}, \ref{fig:md5_connections_3}, and now \ref{fig:md5_connections_4}, this connects that original md5 hash to this large cluster of other indicators of compromise described in two different threat reports. This shows a specific example of using the graph database to connect indicators of compromise together indirectly using network connections. }
    \label{fig:md5_connections_4}
\end{figure*}

\subsection{Visual Analysis: md5 Malware Hash}
\label{section:results_visual_md5_hash}

In order to show a specific malware hash subgraph of the database, we will perform a query to the Neo4j database for the md5 hash \textbf{84c82835a5d21bbcf75a61706d8ab549}. The Cypher language query used to search for this single md5 node is:

\noindent
\texttt{MATCH p=(find:node\_md5 \{name: `84c82835a5d21bbcf75a61706d8ab549'\}) RETURN p}

In order to illustrate the utility of the database, we have selected this hash because it is known to be a hash of a piece of malware. The query results show that there are three unique document nodes that reference this specific hash. Figure \ref{fig:md5_connections_1} shows this connectivity graph. The content of these three document nodes contains useful context information. The lower left hand node document is a cybersecurity blog-style website that is detailing a network and security analysis tool, which used the example of this md5 hash for detecting malware using antivirus software. The lower right hand node document is a manalyzer (manalyzer is an online portable executable static analysis website) report on this md5 hash\footnote{\url{https://manalyzer.org/report/84c82835a5d21bbcf75a61706d8ab549}}. The top node document is a 4chan thread (4chan is an anonymous online discussion forum) in which a user linked to this manalyzer report within a long discussion thread.

Next, we can query the neighbors of those three document nodes - the connectivity graph for this result is shown in Figure \ref{fig:md5_connections_2}. Notably, there is a SHA1 and a SHA256 hash which were both in two of the node documents. These two hashes and the md5 hash are all checksums of the same file. 

Next, we expand the neighbor relationships for these two SHA1 and SHA256 hashes that were both contained in the two threat report nodes in order to see if there are relevant connections we can investigate further. This neighborhood expansion is shown in Figure \ref{fig:md5_connections_3}. We see that there are two node documents that are linked to the SHA256 hash. These two nodes are threat reports. One of these threat reports is also linked to the earlier manalyzer report node by a common filename. However, this filename is simply the command prompt executable (\texttt{cmd.exe}). These two nodes represent slightly different versions of a threat report titled \textit{The Lazarus Constellation}, authored by Avisa Partners. This specific malware hash is associated with the WannaCry ransomware. WannaCry is ransomware \cite{8323680, 9329407, 8702049, 8260673, 8323681} that propagated across the world in 2017 targeting computers running Windows OS primarily using an exploit known as EternalBlue. Figure \ref{fig:md5_connections_4} shows the neighborhood of connections that the \textit{The Lazarus Constellation} threat reports contain, which includes a large number of hashes, CVEs, domains, and filenames that are all associated with this APT group. This malware hash example shows how the graph database can retrieve relevant information for an IOC. Here we were able to find two other hashes of the same file, a place where the hash was mentioned on a 4chan discussion board, and finally threat reports which give the larger context of why these hashes are relevant. 

\begin{figure*}[h!]
    \centering
    \includegraphics[width=0.8\textwidth]{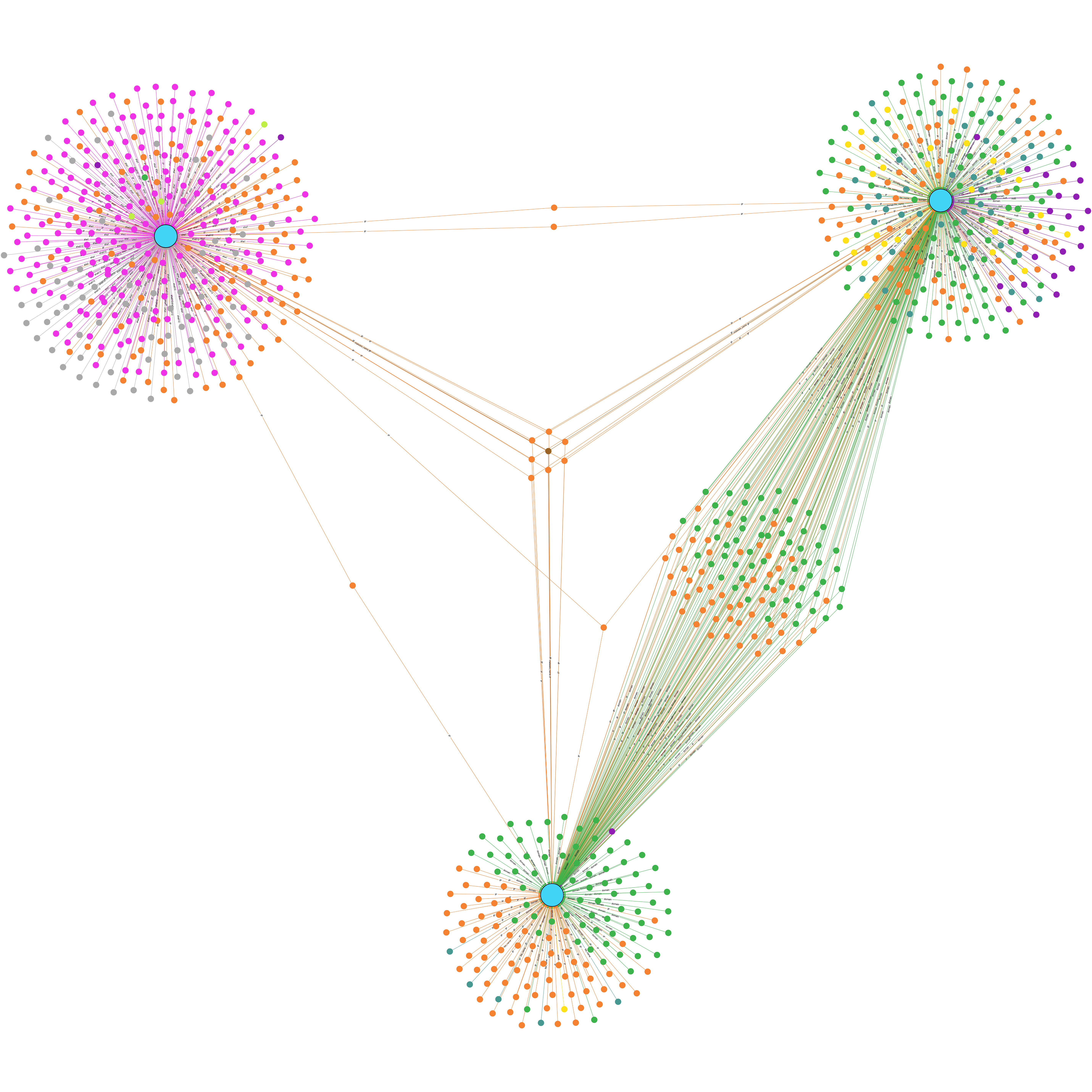}
    \vspace*{-7mm}
    \caption{This is the degree 2 neighborhood halo of connected nodes to a known IP address associated with the Qakbot malware \textbf{89.101.97.139}, where the node representing the IP of interesting is the orange degree 3 node near the bottom document node. This halo of connections is very dense, in particular because all three open source nodes are very high degree, meaning that a large number of datatypes were extracted from those documents when the database was formed. Notably, there are a number of shared datatype nodes, besides the original malware IP address - these include \emph{potential IOCs} of domains and other IP addresses. This type of search scenario illustrates how a single IP address search can potentially retrieve a large collection of related pieces of information.  }
    \label{fig:IP}
\end{figure*}

\subsection{Visual analysis: Qakbot IP address}
\label{section:results_visual_IP_address}

Next we will consider the connectivity graph around a known malicious IP address \textbf{89.101.97.139}, which is the IP address for a command and control server for the malware known as Qakbot from approximately 2021 to 2022. The graph database connections showing where this IP address was mentioned in the database are shown in Figure \ref{fig:IP}. There were three document nodes which mentioned this IP address. Two of these documents were from Github repositories where known IOCs on Qakbot were published, and the third document is from a pastebin page which also posted Qakbot IOCs. Interestingly, there are several IP addresses that are common to all, or a subset, of these documents. Note that the malware name (brown colored degree three node near the center of the graph) was the name Qakbot. This graph shows the relevant context around this IP address - namely that it is a command and control server for Qakbot and the connectivity graph shows many additional IOCs that should be monitored in association with Qakbot. Importantly, these graph connections take advantage of multiple (separated) data sources - where if we were to have read only a single one of these documents we would not have the aggregated data which includes many more IP addresses and hash IOCs. It is clear from this graph that this specific IP address is not a lone command and control server for Qakbot, rather it is one of many IP addresses and domains that are connected to this malware. Figure \ref{fig:large_graph_renderings} in the Appendix \ref{section:appendix_large_graphs} shows significantly larger graph renderings for the degree 3 and 4 connections in the database from this IP address node. 

\subsection{Visual Analysis: Known Exploited CVE}
\label{section:results_visual_CVE_exploited}
\begin{figure*}[h]
    \centering
    \vspace*{-22mm}
    \includegraphics[width=0.8\textwidth]{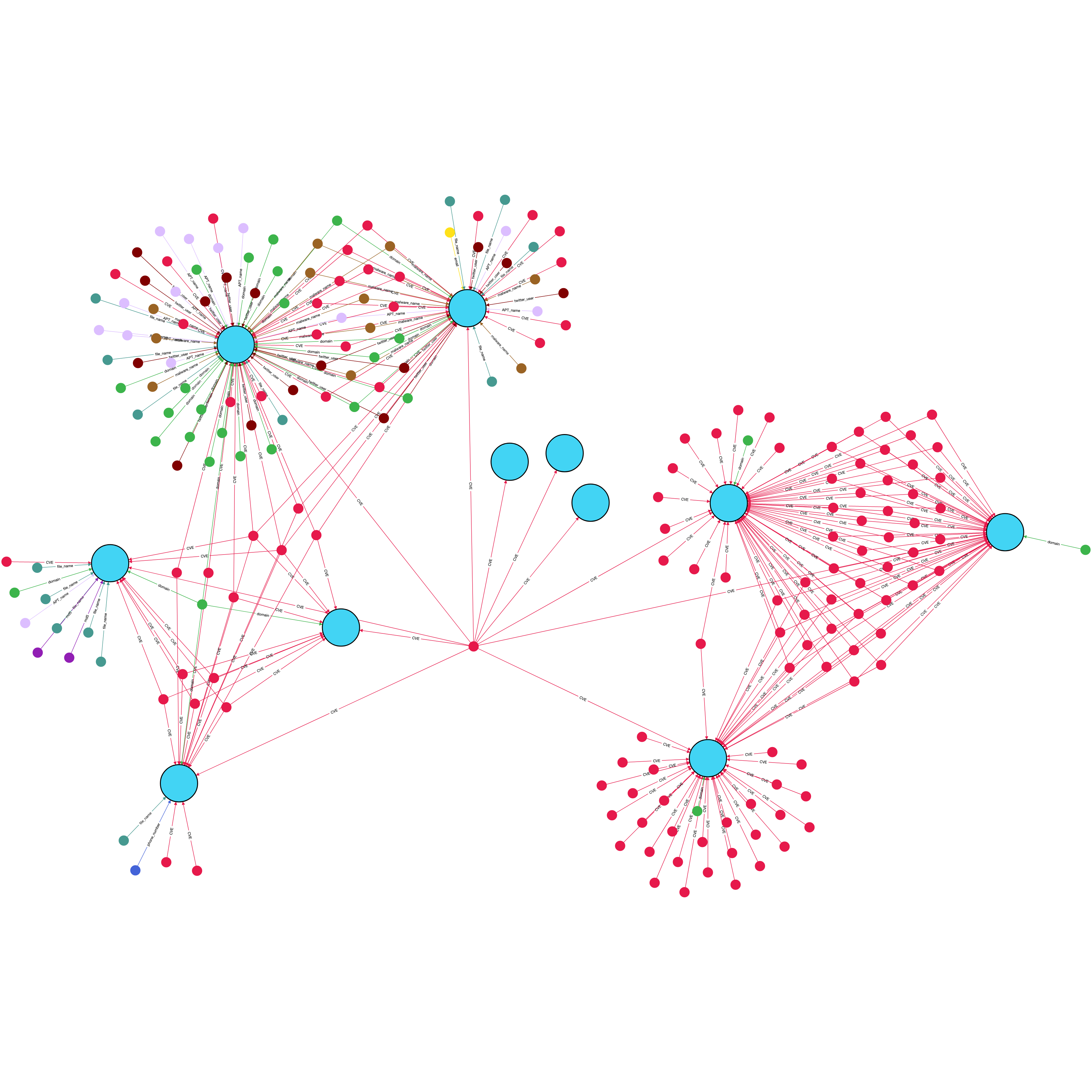}
    \vspace*{-27mm}
    \caption{Shown here are all nodes and edges connected to the node for CVE-2014-4404 up to two degrees away. The degree 11 red node approximately in the center of the graph is the CVE-2014-4404 node. The 11 light blue nodes denote 11 open source documents which mention CVE-2014-4404 - all other extracted datatypes from those 11 documents are also shown. Notably, several of these documents mention numerous other CVE numbers, suggesting they could be security bulletins or repositories of some type. Other datatypes in these associated documents include hash checksums, potential malware names, APT group names, and a potential email address. All of these linked datatypes could be IOCs that are linked to CVE-2014-4404. }
    \label{fig:CVE_connections}
\end{figure*}

Here we show the connectivity graph for all neighboring nodes up to degree 2 away from a CVE that is known to be exploited in the wild\footnote{\url{https://www.cisa.gov/known-exploited-vulnerabilities-catalog}}; CVE-2014-4404. CVE-2014-4404 is a remote code execution exploit on Apple OS X, earlier iOS versions and Apple TV. The neighborhood graph is shown in Figure \ref{fig:CVE_connections}. The neighboring nodes connected to this specific node are showing the halo of data that is connected to this CVE. In particular, these neighboring nodes show filenames, other CVEs, twitter usernames, domain names, hashes etc., that were mentioned along with this CVE in various documents. Two of these nodes were threat reports which described the state of various cyberattacks in 2021 - therefore they broadly discussed a number of different exploits, APT groups, twitter usernames, and malware families with CVE-2014-4404 being one of the more exploited Apple software CVEs. As one would expect, the CTC tool labelled each of these text documents that are in this halo as being cybersecurity related. This ML metadata can be used to filter for a subgraph of the graph database which contains only cybersecurity related (English) natural language. 

\subsection{Malware hash files: code and resource re-use}
\label{section:results_malware_PE_shared_resource}
The static analysis scan hash digest \cite{hahn2014robust} nodes that are formed in Neo4j are comprised of hashes of extracted files, functions, or strings contained within the malware (as well as the malware itself). What these additional hashes allow within the graph database is detection of \emph{shared} code components, for example shared header files, strings, or functions. Of course some of these may not be incredibly useful indicators - for example they may be standard Portable Executable manifest files. However, if there are unique code artifacts in some subset of antivirus file scans which are known to be malicious, then those could serve as indicators of shared code re-use from a group. These could even serve as malware signatures for antivirus products. The graph database is well suited for investigating these shared resource hashes because it is easy to find where a set of antivirus scans have a shared resource hash node. 

Here we provide a specific example - that is a SHA256 hash which we can search for in the Neo4j graph database with the Cypher syntax of:

\noindent
\texttt{MATCH p=(find:node\_sha256 \{name: '84f7c54dc015637a28f06867607c2e0b \newline dd225d10debb1390ff212d91cd2d042b'\}) RETURN p}

Using FileScan\footnote{\url{https://www.filescan.io}}, which contains references to this hash, we can find that this is a hash of the following English ASCII text: \texttt{BundleInstall BundleInstall}. Note that this representation is not necessarily capturing the full hexadecimal data present in the data segment. The terms bundle install suggest that the source code language is Ruby, and this text artifact could be a result of packaging Ruby source code into a portable executable. 

\begin{figure}[h!]
    \centering
    \includegraphics[width=0.5\textwidth]{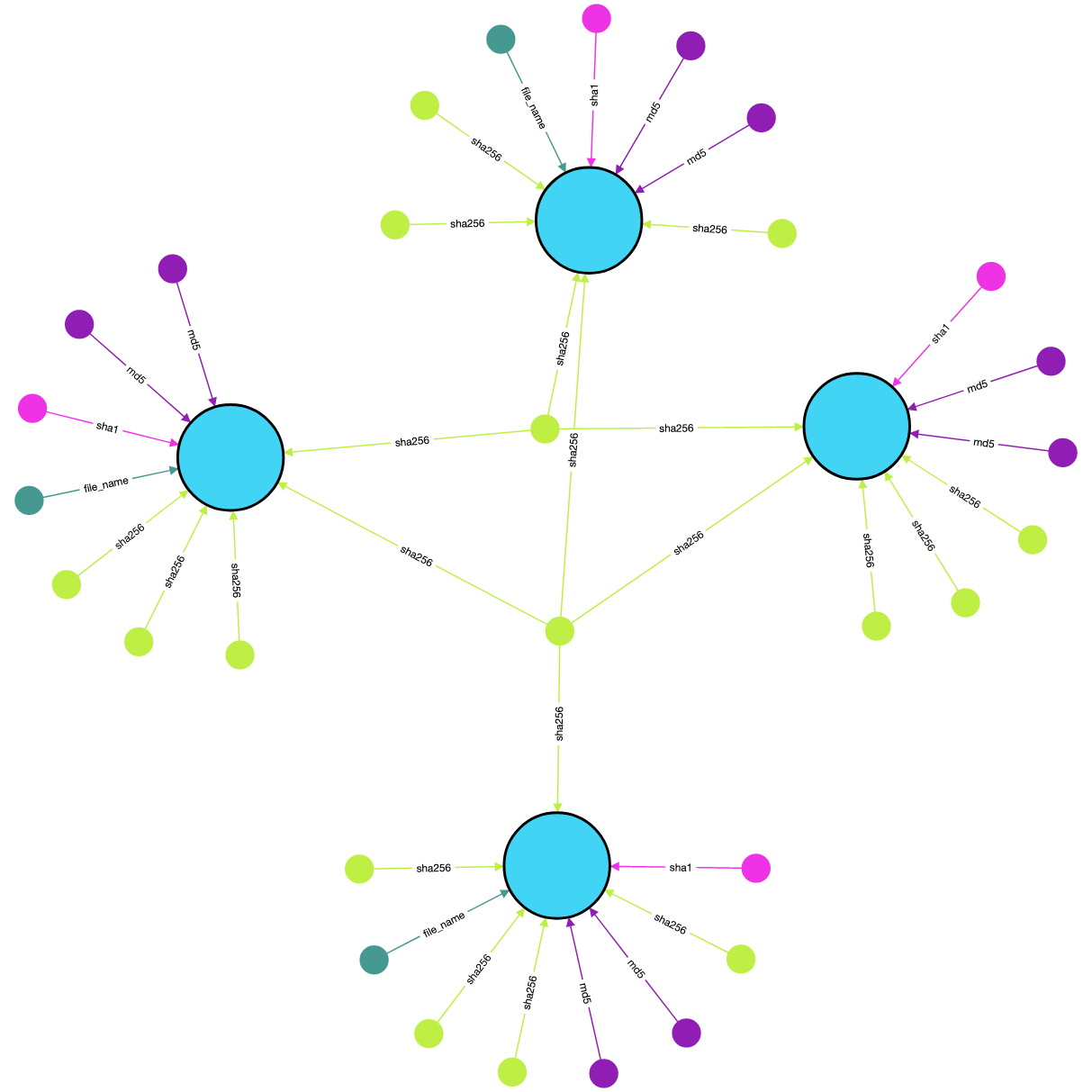}
    \caption{This is a graph rendering of the neighboring nodes connected to the sha256 node hash of interest, which is the degree four green-yellow node in the center of the graph. The neighboring connections of these document nodes have also been displayed in order to determine if there are any other connections of interest. 
    Each of the document nodes are antivirus scans of portable executables (PEs). The degree 1 expanded neighbors show that there is actually another sha256 hash node that is shared by three out of the four document nodes, and there are no other edges connecting the associated hashes and file names. That sha256 hash is a standard manifest for creating PEs and is common to a large number of PE samples (both benign-ware and malware) and is therefore not unique enough to attribute a meaningful connection. }
    \label{fig:PE_neo4j_resource}
\end{figure}

Within the current Neo4j database, this SHA256 hash node has a degree of 5 (meaning that is referenced in 5 node documents). Two of these nodes are actually effectively duplicates - they are scans of the same file, meaning that they have the same file name and associated extracted hashes (such as the hashes of contained resources in the Portable Executable (PE) file), but were scanned at different times and therefore have slightly different data so the direct de-duplication did not remove one of them. Therefore as with the other Neo4j figures in order to reduce clutter, one of these duplicates is removed in the displayed figures. The graph renderings of these nodes and their degree 1 neighborhood connections are shown in Figures \ref{fig:PE_neo4j_resource}. 

For the bottom graph in Figure \ref{fig:PE_neo4j_resource} we can examine what some of the important information is that the nodes have, such as the malware filename and the hash of the file. These four nodes are known malware samples (this statement is based on the high proportion of antivirus scans indicating malware) and additional information on them is available on VirusTotal. 

\noindent
\textbf{Top document node} in Figure \ref{fig:PE_neo4j_resource}: the filename is \texttt{rkinstaller.exe} and the full sha256 hash is given in the virustotal link.\footnote{\url{https://www.virustotal.com/gui/file/5577ce9aa4e4ec2735247c5769f0e84db599825f2d95159b0102f3b30e80b6bb/details}} 

\noindent
\textbf{Right document node} in Figure \ref{fig:PE_neo4j_resource}: no associated filename and the full sha256 hash is given in the virustotal link.\footnote{\url{https://www.virustotal.com/gui/file/06f11f4a555a4891c93f13f82dc06e8bcedda2a71c8a5e6aa5c18da871f41238/details}} 

\noindent
\textbf{Bottom document node} in Figure \ref{fig:PE_neo4j_resource}: the filename is \texttt{rkinstaller364.exe} and the full sha256 hash is given in the virustotal link.\footnote{\url{https://www.virustotal.com/gui/file/f8d11b1e3e027355a11163049b530de4fd67183abd08a691d5d18744653ef575/details}}

\noindent
\textbf{Left document node} in Figure \ref{fig:PE_neo4j_resource}: the filename is \texttt{poinstaller257.exe} and the full sha256 hash is given in the virustotal link.\footnote{\url{https://www.virustotal.com/gui/file/f3efcfc7121f2348deb6f3b5ffde60878d978c25281e67defdc288feaef8b38c/details}}

\begin{figure}[h!]
    \centering
    \includegraphics[width=0.49\textwidth]{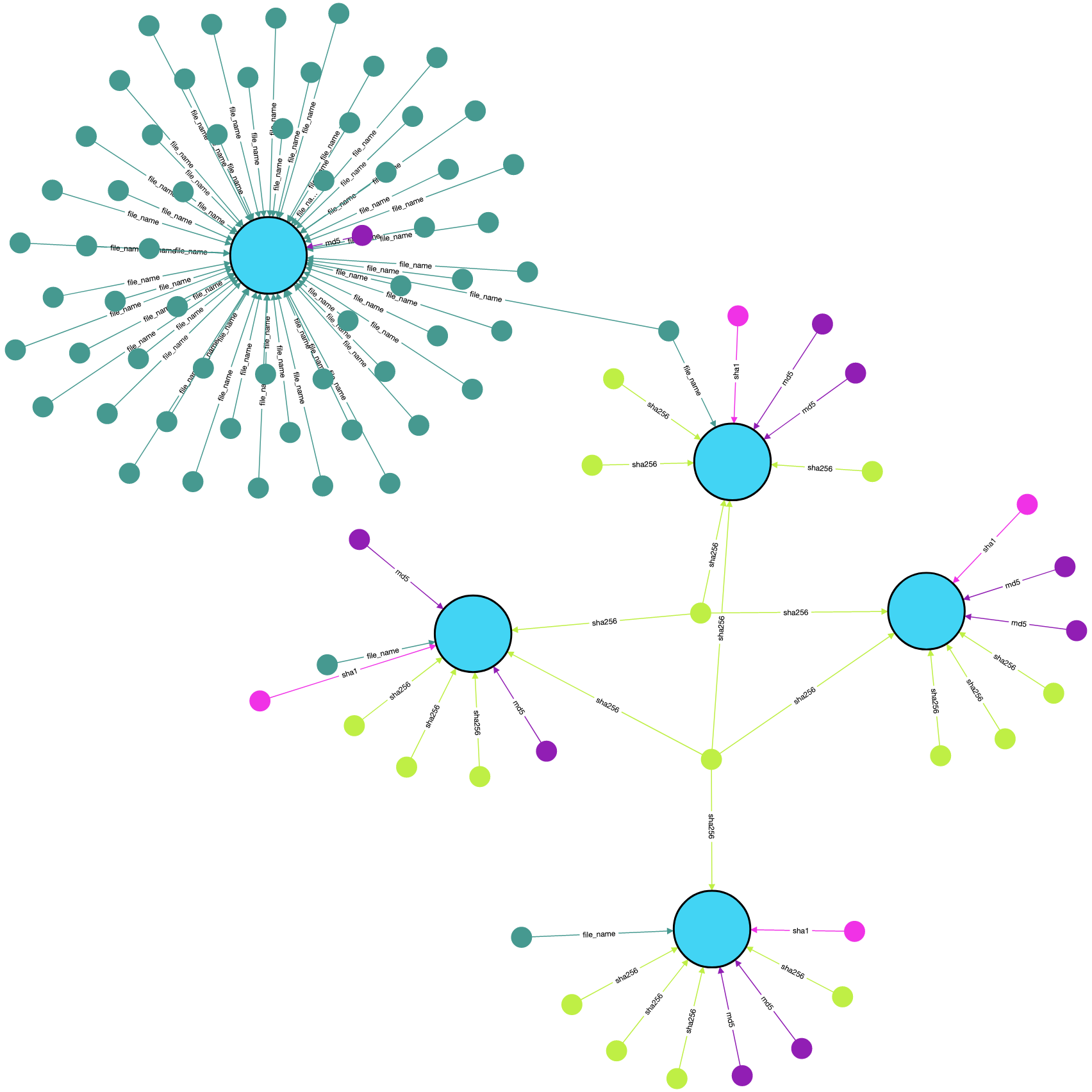}
    \caption{This graph is a continuation of Figure \ref{fig:PE_neo4j_resource} where further mentions of the filenames and hashes from the antivirus scans are displayed. One of the sha256 nodes in the bottom document node is also contained in many other node documents, however it is a hash of a standard dynamic link library (DLL) manifest file and therefore its neighbors are not included in this figure to reduce visual clutter. The only other node that was referenced outside of this small connected portion of the graph was the filename in the top document node, which was \texttt{rkinstaller.exe}. This filename was also mentioned in a site that listed a large number of filenames known to be malware - that node document along with the halo of its connected potential IOCs (most of which are other filenames) are shown in the upper left hand portion of the figure. }
    \label{fig:PE_neo4j_resource_2}
\end{figure}

The notable observation from this data is that this shared portable executable resource hash is a reasonably unique artifact (meaning that it is not a commonly re-used ASCII text segment in PE development) and was found across a small subset of malware samples, some of which also have other shared characteristics such as similar (or identical) file names. This suggests, with reasonable confidence, that the development of these pieces of malware is linked in a meaningful way, for example the same developer could have created these portable executables. This demonstrates where the graph database construction of OSint allows a user to link together pieces of information in order to group together seemingly unconnected documents and other potential indicators. 

Portable executable visualizations for three other pieces of known malware which contain this specific PE resource sha256 hash are shown in Figure \ref{fig:PE_visuals} in Appendix \ref{section:appendix_PE_visual}.

\subsection{CVE Degree and CVSS Score}
\label{section:results_CVE_degrees}

\begin{figure*}[h!]
    \centering
    \includegraphics[width=0.49\textwidth]{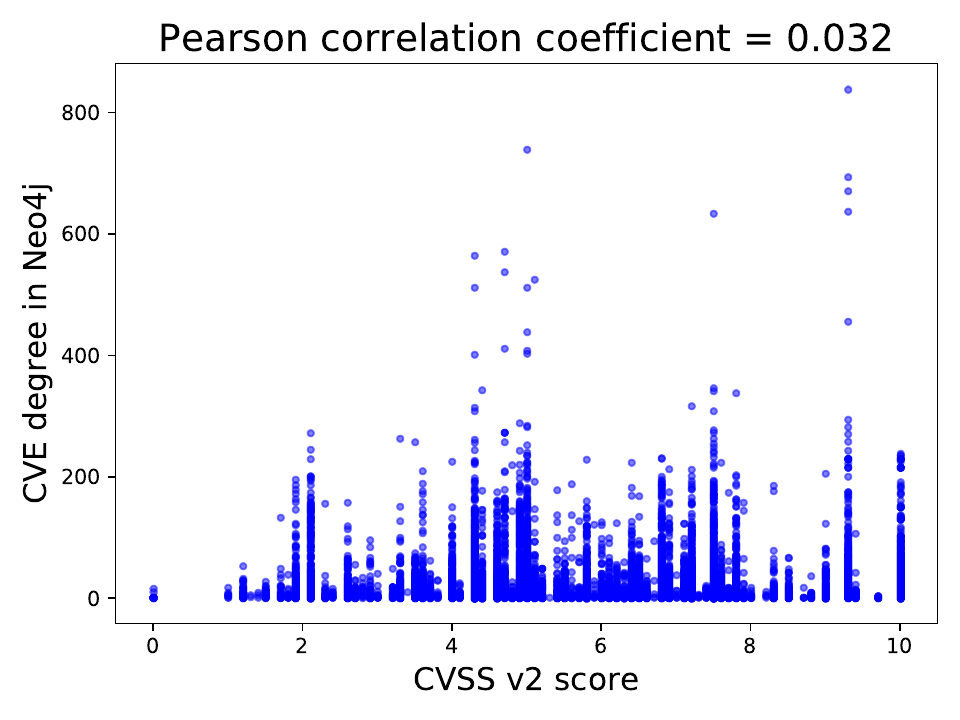}
    \includegraphics[width=0.49\textwidth]{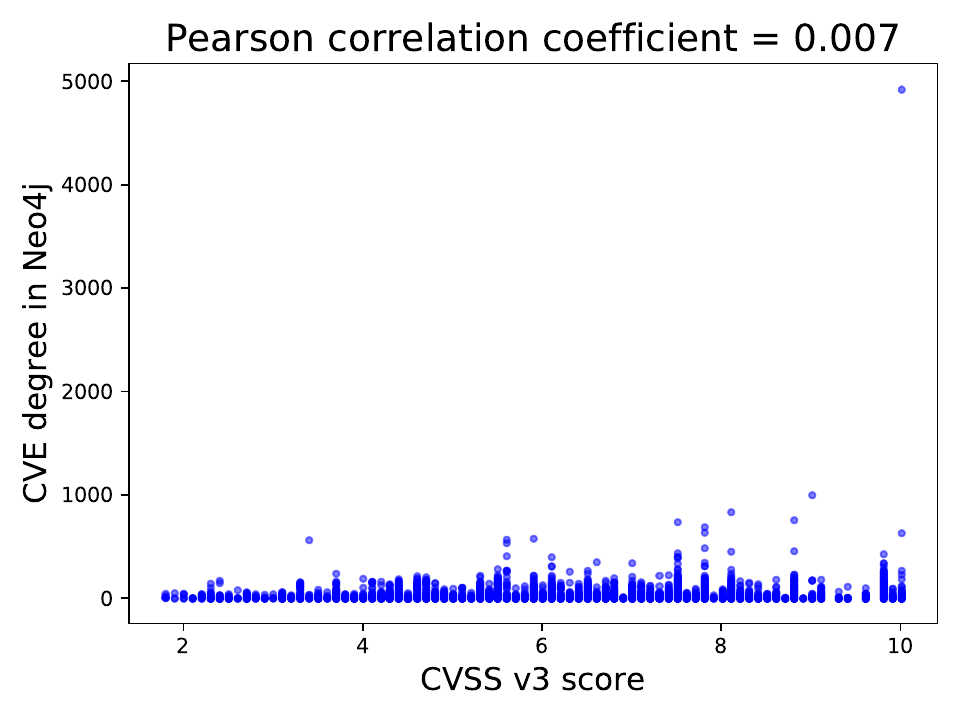}
    \caption{CVSS score (x-axis) version 2 (left) and version 3 (right) vs CVE node degree in the Neo4j indicator database (y-axis). Right hand figure contains $97489$ datapoints, and the left hand figure contains $150940$ datapoints. The pearson correlation coefficient for each dataset is shown in the plot titles. The outlier CVE node in the right hand plot, which has a degree of $4926$, is \texttt{CVE-2021-44228} (also known as Log4j). }
    \label{fig:CVE_CVSS_score_vs_degree}
\end{figure*}

\begin{figure*}[h!]
    \centering
    \includegraphics[width=0.49\textwidth]{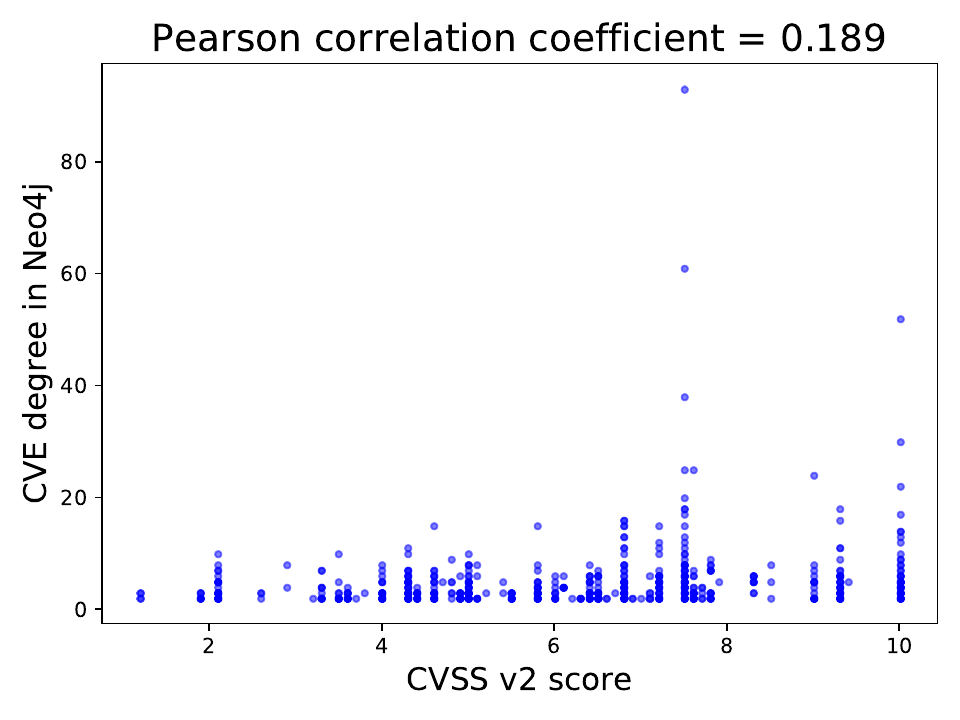}
    \includegraphics[width=0.49\textwidth]{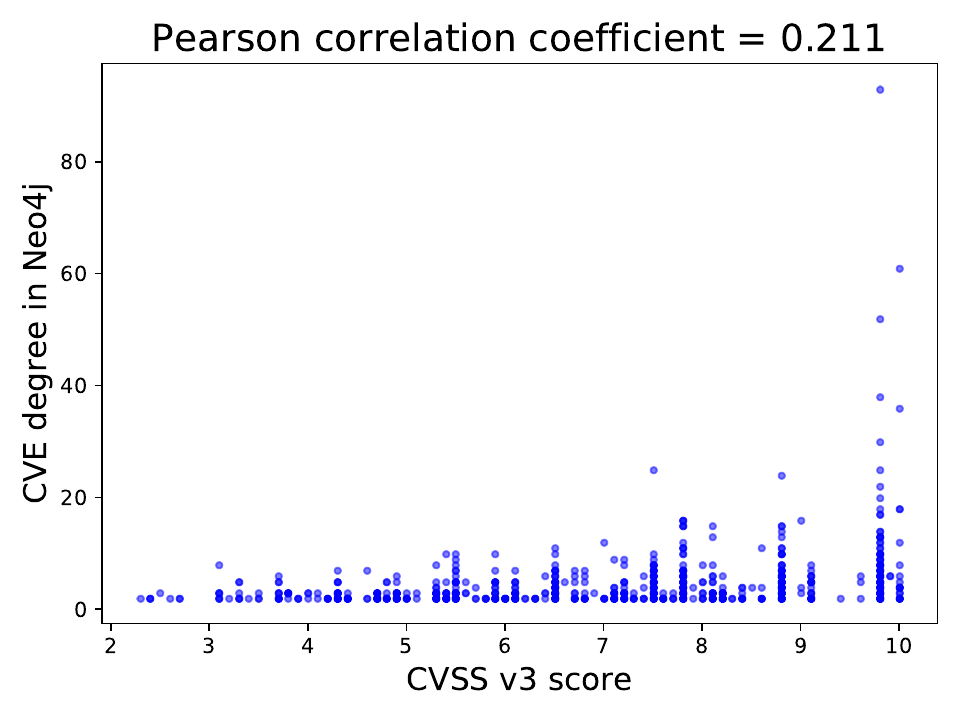}
    \caption{CVSS score (x-axis) version 2 (left) and version 3 (right) vs CVE node degree in the Neo4j indicator database (y-axis). The datapoints plotted are based on the degree of the CVE nodes connected to a \emph{subset} of the document nodes which are more reputable for the domain of cybersecurity (e.g., threat reports), and CVE nodes with degree $1$ are not considered. Additionally, only CVEs which were released in the time period of the web scraping upon which the graph database is built, are plotted in order to remove temporal bias which have existed in Figure \ref{fig:CVE_CVSS_score_vs_degree}. The pearson correlation coefficient is shown in the plot titles. Right hand figure contains $1898$ datapoints, and the left hand figure contains $1666$ datapoints.  }
    \label{fig:CVE_degree_CVSS_restricted}
\end{figure*}

Because this database is constructed in a largely unsupervised manner - i.e., pattern matches are automatically generated and new data is added to the database without human review - a natural question that arises is whether the graph structure of the data represents the real world properties of the vulnerabilities or potential IOCs. An easy example of this that we can numerically compute is the degree of CVE nodes in the database (which corresponds to how many times that CVE was mentioned in the text from the different data sources) and compare that against the common vulnerability scoring system (CVSS) scores of CVEs. Ideally, this type of ground-truth correlation analysis could be applied to other objects in the graph database, but CVEs and CVSS scores are a very direct and natural correlation to study here. CVSS scores are intended to approximately represent the overall \emph{severity} of a security vulnerability \cite{cheng2012aggregating, scarfone2009analysis, khazaei2016automatic, gallon2011using} where a CVSS score of $0$ is the lowest severity and $10$ is the maximum severity. There are two CVSS score versions that we will compare - CVSS version 2 and version 3.\footnote{Note that there is also a version 4 scale which is intended to have even more accurate vulnerability severity measurement, although we do not compare this version} Version 3 being the newer of the two CVSS scoring methods, which is intended to be a more accurate rating scale for modern cybersecurity threats. We specifically use the CVSS \emph{base scores} (\textbf{CVSS-B v2} and  \textbf{CVSS-B v3}), which do not form a complete severity picture of the CVE (for example these do not include temporal factors that will change over time). The base scores are intended to represent the foundational severity score of the vulnerability. Importantly, all CVSS versions are in some way approximating the notion of vulnerability severity, and are not perfect scoring systems. However, they are intended to be operationally useful for understanding, mitigating, and prioritizing resources to handle vulnerabilities. In this sense, these scores are excellent test cases of real-world data to compare our open source graph database to.

The relevant question is whether there exists a relationship of increasing degree of the CVE nodes in the graph database with respect to CVE CVSS score. Intuitively, if the severity of a CVE corresponds to how frequently that CVE is mentioned in social media, news, and threat reports, then higher CVSS scores will correspond to higher degree CVE nodes in the Neo4j graph database. The CVSS scores are retrieved from the National Institute of Standards and Technology National Vulnerability Database (NIST NVD) dataset.\footnote{\url{https://nvd.nist.gov}} To this end, we compute the Pearson correlation coefficient between CVE node degrees and their CVSS score using scipy in Python 3 \cite{2020SciPy-NMeth, student1908probable, kowalski1972effects, benesty2009pearson}. The Pearson correlation coefficient is a simple statistic that describes how linearly correlated two datasets are, in this case the CVE node degree within the Neo4j database and the CVSS score of that CVE. A correlation coefficient of $0$ indicates no correlation, $1$ indicates a strong positive linear relationship, and a $-1$ indicates a strong negative linear relationship in the data. The motivation for measuring the correlation coefficient is that it would be reasonable to hypothesize that more severe CVE scores correspond to more discussion in open source venues on the internet, and in threat reports. Some CVEs do not have a (measured, and in the NVD database) version 2 or a version 3 score, and therefore are not plotted in this dataset. Some CVEs do not have a v3 score, but do have a v2 score, or conversely have a v2 score but no v3 score. This means that the correlation between the v2 and v3 data are not shown on exactly the same set of CVE nodes. This also means that there is a certain temporal correlation in the data, namely that CVEs that do have v2 scores are older, whereas CVEs that have v3 scores are newer.

Figure \ref{fig:CVE_CVSS_score_vs_degree} plots all CVE CVSS scores against CVE node degrees in the Neo4j graph database, which shows there is not a strong positive or linear correlation between the CVE node degrees and CVSS scores. This is notable because it shows that across all of the cybersecurity mentioned natural text that was gathered, there is not a strong correlation between the rate of CVE mentions (e.g. CVE popularity) and CVSS scores. However, it could be the case that more focused cybersecurity documents have a higher CVE CVSS score and Neo4j node degree correlation. It could also be the case there is not a strong signal of correlation for CVE nodes which are not commonly mentioned in news, threat reports, and cybersecurity bulletins - and therefore only considering nodes which have at least a degree of $2$ could remove some noise in the dataset. Lastly, due to popularity of new CVEs, necessarily the web scraping will have a temporal bias for the times during which the spiders were operating. 

Figure \ref{fig:CVE_degree_CVSS_restricted} shows the correlation plot for CVE node degrees vs CVSS score\footnote{Note that all of the reported data ends in 2022 and does not include any subsequent years. }, but the Neo4j node degrees are computed using a restricted set of document sources namely reputable blogs, information sites, and threat reports - specifically \texttt{fireeye}\footnote{\url{https://www.trellix.com/en-us/about/newsroom/stories/threat-labs.html}}, all threat reports, \texttt{proofpoint}\footnote{\url{https://www.proofpoint.com/us/blog/threat-insight}}, \texttt{exploitdb}\footnote{\url{https://www.exploit-db.com/}}, \texttt{the Hackernews}\footnote{\url{https://thehackernews.com/}}. Additionally, CVE nodes with degree $1$ are not included in this computation since low degree nodes do not necessarily provide a strong signal in regards to how referenced that CVE is (degree one nodes can be interpreted as a type of finite sampling noise in the dataset). Lastly, the data in Figure \ref{fig:CVE_degree_CVSS_restricted} are restricted to the years during which the OSint crawler system was operational \cite{9717379} because there is an inherent temporal bias in regards to what web pages are mentioned and scraped. Because only a subset of the CVEs had CVSS v3 scores, there were fewer data points available for those plots. In Figure \ref{fig:CVE_degree_CVSS_restricted} we observe that there is a weak to median linear positive correlation between CVSS score and Neo4j degree for the documents that are cybersecurity domain focused. Interestingly, there is a slightly weaker Pearson correlation ($0.196$) for CVSS version 2 compared to version 3 ($0.23$). Figure~\ref{fig:CVE_degree_CVSS_restricted} shows that taking into account the source of the data, as well as temporal bias (by only considering the CVEs that were released during the graph database construction), and not considering degree $1$ nodes shows that there is a weak linear CVSS and CVE node degree correlation, as opposed to these correlations shown in the entire dataset in Figure~\ref{fig:CVE_CVSS_score_vs_degree}. This shows that there is some measurable correlation between vulnerability severity and open source cybersecurity text occurrence frequency. There are two potential reasons why the CVSS score - node degree correlations are only weakly linear in Figure \ref{fig:CVE_degree_CVSS_restricted}:

\begin{enumerate}
    \item The inherent bias present in the documents from the web crawlers - i.e., a CVE could be more commonly discussed on social media or news sites because of a reason other than its significance for cybersecurity. This reason seems to be the most prevalent due to the presence of social media and news content. In essence, the frequency of mentions of a CVE are more related to what catches attention than severity of the vulnerability. Examples of this bias include popular software being discussed more than less known software. Building on these findings of Figure~\ref{fig:CVE_CVSS_score_vs_degree} and \ref{fig:CVE_degree_CVSS_restricted}, this potential of bias could be corrected for in future work by defining and quantifying a measure of software ubiquity, and then correcting the node degree based on a popularity measure. 
    \item The CVSS base score does not perfectly reflect the real world severity of a given CVE. 
\end{enumerate}

Note that the observation of more severe CVE CVSS scores not necessarily correlating with popularity has been observed before such as in ref.~\cite{spring2019prioritizing}. 

\begin{table*}[h!]
\centering
\begin{tabular}{||c || p{1.7cm} || p{6.5cm} | p{1.7cm} | p{0.8cm} | p{0.8cm} ||} 
 \hline
 CVE ID & CVE node Page Rank score in Neo4j & Vulnerability name and description & In CISA known exploited vulnerability catalog & CVSS base score v2 & CVSS base score v3 \\ [0.5ex] 
 \hline\hline
 CVE-2021-44228 & 758.1 & Apache Log4j2 Remote Code Execution Vulnerability & Yes & 9.2 & 10.0 \\ 
 \hline
 CVE-2021-45046 & 113.06 & It was found that the fix to address CVE-2021-44228 in Apache Log4j 2.15.0 was incomplete in certain non-default configurations. & Yes & 5.1 & 9.0 \\
  \hline
 CVE-2021-34527 & 106.21 & ``PrintNightmare'' - Microsoft Windows Print Spooler Remote Code Execution Vulnerability & Yes & 9.0 & 8.8\\
  \hline
 CVE-2017-11882 & 95.64 & Microsoft Office memory corruption vulnerability & Yes & 9.3 & 7.8 \\
  \hline
 CVE-2012-0158 & 86.71 & Microsoft MSCOMCTL.OCX Remote Code Execution Vulnerability & Yes & 9.3 & N/A \\ 
 \hline
 CVE-2014-0160 & 81.13 & OpenSSL Information Disclosure Vulnerability. ``heartbleed'' & Yes & 5.0 & 7.5 \\
 \hline
 CVE-2021-34481 & 73.47 & Windows Print Spooler Elevation of Privilege Vulnerability & No & 4.6 & 7.8 \\ 
 \hline
 CVE-2021-45105 & 70.92 & Apache Log4j2 versions 2.0-alpha1 through 2.16.0 (excluding 2.12.3 and 2.3.1) did not protect from uncontrolled recursion from self-referential lookups. & No & 4.3 & 5.9 \\ 
 \hline
 CVE-2021-1675 & 64.43 & Microsoft Windows Print Spooler Remote Code Execution Vulnerability & Yes & 9.3 & 8.8 \\ 
 \hline
 CVE-2021-40444 & 59.45 &  Microsoft MSHTML Remote Code Execution Vulnerability  & Yes & 6.8 & 7.8 \\ 
 \hline
 CVE-2017-0199 & 56.32 & Microsoft Office/WordPad Remote Code Execution Vulnerability with Windows API & Yes & 9.3 & 7.8 \\ 
 \hline
 \noalign{\vskip 1mm}
\end{tabular}
\caption{The top 11 most referenced CVEs ranked by their PageRank score in the Neo4j database in descending order. The PageRank computation was performed on all edges, regardless of type, in the database. PageRank scores are rounded to two decimal places. Note that the Cybersecurity \& Infrastructure Security Agency (CISA) known exploits catalog is continuously updated - this information is correct at the time this paper is written, but may change in the future. The PageRank score ranking is the most important part of the table, but the numerical page rank scores do approximately correspond to the degree of popularity and relevance of the CVE id. For example, it is clear that in the current dataset CVE-2021-44228 is significantly more referenced than every other CVE in the graph database. The intention of including the CVSS scores and whether the vulnerability is currently in the catalog, as of January 1 2025, is to give additional context on the severity of the vulnerability. }
\label{table:page_rank_CVE}
\end{table*}

\begin{figure*}[h!]
    \centering
    \includegraphics[width=0.42\textwidth]{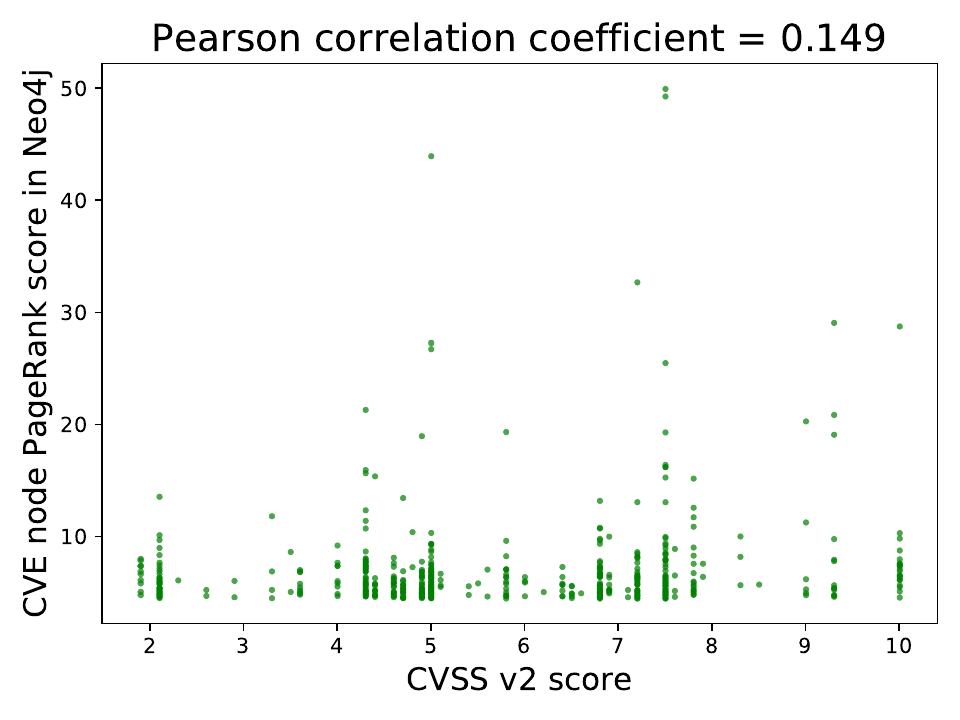}
    \includegraphics[width=0.42\textwidth]{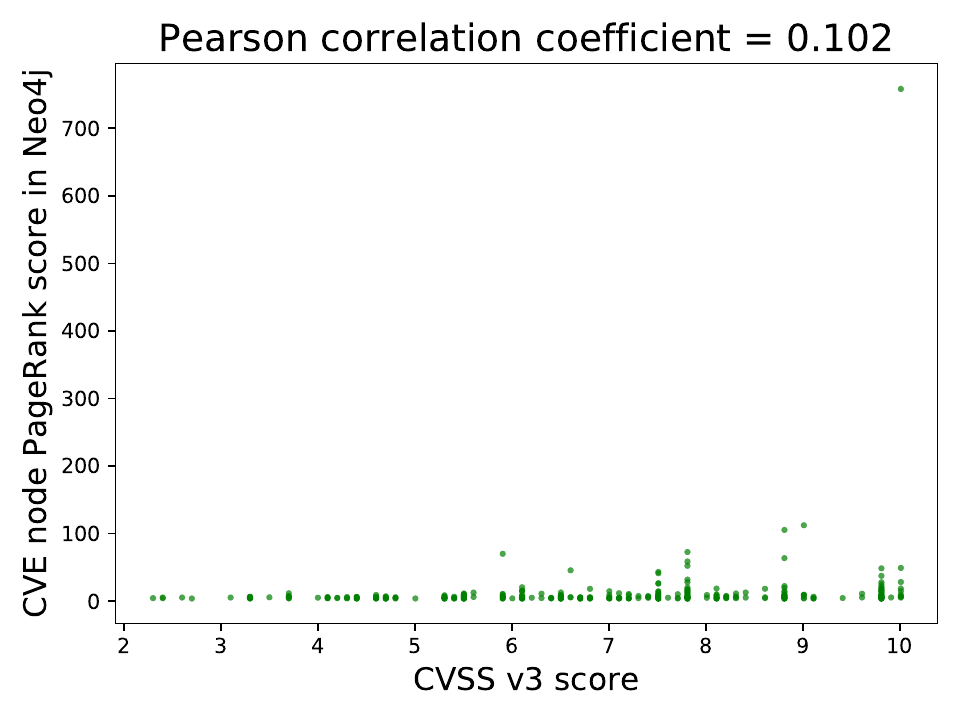}
    \caption{CVSS score (x-axis) version 2 (left) and version 3 (right) vs CVE node PageRank score in the Neo4j indicator database (y-axis). Right hand figure contains $715$ datapoints, and the left hand figure contains $567$ datapoints. The pearson correlation coefficient is shown in the plot titles.  }
    \label{fig:CVE_CVSS_vs_PageRank}
\end{figure*}

\subsection{CVE Page Rank}
\label{section:CVE_page_rank}
Neo4j has a library called \emph{Graph Data Science} (GDS) which contains various graph algorithms that can be executed on a graph that is stored in Neo4j. These graphs algorithms could be useful for identifying patterns and clusters in this specific potential IOC graph database. As a simple example of how the structure of the graph yields node rankings based only on the reference to potential IOCs and vulnerability IDs, Table \ref{table:page_rank_CVE} details the top 11 highest ranked CVEs in the (undirected) graph database according to the PageRank algorithm. PageRank \cite{brin1998anatomy, gleich2015pagerank} is an algorithm, originally designed for search engine ranking, which can be applied to any network data structure to determine which nodes are the most influential and referenced (here a reference is simply an edge in the network). All parameters were set to default for the PageRank computation with the exception of \texttt{maxIterations} which was set to $300$ and \texttt{dampingFactor} which was set to $0.75$ instead of the typical $0.85$. The reasoning for selecting a smaller damping factor than what is typically used in search engines is that in this specific graph we are interested in potentially longer range influences on the relevance of nodes. 

The PageRank scores in Table \ref{table:page_rank_CVE} clearly show that the most referenced CVE nodes in the graph database are high impact and well known CVEs including the heartbleed, Log4j, and PrintNightmare vulnerabilities. All of these CVEs not only are high impact and typically rank high on CVSS scores, but they are also notable because of how widespread their discussion was throughout the various social media channels, in large part because of the wide user base of the software these vulnerabilities exploit, that this graph database is constructed from. 

Another natural question that arises from the PageRank scores, is how those scores relate to the CVE CVSS score. Table \ref{table:page_rank_CVE} shows that there does seem to be at least some correlation, where the high PageRank score CVEs have high CVSS scores. Similar to the degree correlation plots in Figures \ref{fig:CVE_CVSS_score_vs_degree} and \ref{fig:CVE_degree_CVSS_restricted}, in Figure~\ref{fig:CVE_CVSS_vs_PageRank} CVE node PageRank scores are plotted against CVSS scores. Because the majority of the OSint crawler data is temporally biased towards the more recent years of data gathering, the datapoints plotted in Figure \ref{fig:CVE_CVSS_vs_PageRank} are restricted to the years during which the OSint crawler system was operational see \cite{9717379} for more details, the same as in Figure \ref{fig:CVE_degree_CVSS_restricted}. Additionally, in order to filter only for nodes which have a robust distribution of neighbors in the graph, only points with a PageRank score of $4.5$ or over are plotted. Figure \ref{fig:CVE_CVSS_vs_PageRank} shows that there is a low to medium linear correlation for the relevance, i.e. the PageRank score, against the CVE CVSS scores. This once again shows that there is some correlation between the graph structure in the database and the real world proxy vulnerability severity scores of CVSS v2 and v3.

\section{Discussion and Conclusion}
\label{section:discussion}

We have described and presented quantitative metrics of a Neo4j graph database of cybersecurity relevant data from open source text sources. We have highlighted several key examples within the graph database where known malicious indicators are identified in internet discussions. We also examined whether there exist correlations between CVE score severity and the frequency at which that CVE is mentioned, in the constructed database. Importantly, the Neo4j system that we constructed is very flexible in terms of the types of queries that can be made to the database; either in the form of Cypher language queries, or using a Python library to interact with the database. This means that searching for document entries with complex parameters can be performed using this Neo4j graph database. In this study we focus on visual small scale clear examples where specific indicators were found. There are primarily two technical challenges that still need to be improved in this data gathering and analysis pipeline:

First, there are many instances of \emph{near duplicate} content from the web crawling system. For example, the content of a web page could be slightly altered from day to day (for example if the web page content includes the current date and time); and if the web crawlers end up at that same web page multiple times direct de-duplication will not remove the content for being a near duplicate. Another example of near duplicate content that occurs often is that the web-crawling will catch a social media conversation (for example a Reddit thread) while it is occurring. We want the crawlers to catch this type of conversation because they could save content that is removed at some point in the future, but it also leads to potentially a large number of near duplicates of the same social media thread as it evolves over time. The challenge is in quantifying how close a document is to being a duplicate in order to remove it; computing a distance metric pairwise between all documents can be very computationally intensive. A reasonable solution could be to not remove near duplicates, but instead to take the union of near duplicate documents, and to compute similarity metrics among close clusters of documents (for example determined by their relationship in the Neo4j potential IOC graph) for finding near duplicates.  

Second, reduce the amount of \emph{noise} in the graph database. It is difficult to know a-priori which content is relevant, and therefore in this work we tend to gather more information rather than remove potentially not useful information. This allows the database to be able to catch interesting edge cases and atypical cybersecurity content, but it comes at the cost of increased noise. However, there are some consistent sources of noise which are almost always not relevant which can be manually filtered out. For example the localhost IP address (\texttt{127.0.0.1}) or standard file names from popular programs. Another example of noise in the data set is software version numbers being identified as IP addresses due to the similarity in their format in text. It is not clear how to reduce the noise in the data set uniformly across all document types of potential IOCs, however, machine learning algorithms which more closely identify relevant information in a piece of text could be used to better filter the data. Importantly, the fact that we have many connections between nodes that are ``false positives'' is inherently part of the system that we describe; verifying the existence and validity of each individual potential indicator would be extremely time consuming and in general cannot be automated. A good example is filenames, because there are many portable executables with the same name, but at the stage of open source text scraping we have no way of verifying whether the two filenames are actually referring to the same portable executable. Instead, we focus on ingesting a high volume of information, and parsing that information with efficient rules and heuristics, in order to obtain a broad set of documents and potential indicators. 

The additional metadata that we automatically gather, such as natural language categorization and machine learning cybersecurity topic modeling, are a subject that we did not report in in Section~\ref{section:results} because pivoting through the graph database based on these more ``fuzzy'' metrics results in high levels of noise, and generally finding a large number of nodes that match a general query. For this reason, we instead propose that these types of metadata could mostly be useful as high level document content summaries for a potential user, instead of manually reading through all content in a document. 

There are additional potential indicators and vulnerability tracking ids that could be extracted from natural language text in future work, such as CWE's (Common Weakness Enumeration). 

Aggregating and condensing open source intelligence into a human readable and easily searchable form is an important task. This is especially true due to the scale of the data that is available in the form of social media, news, blogs and threat reports in the cybersecurity space. Here we have presented one possible way to address this problem by parsing and transforming the open source data into a graph structure where each document can be associated with potential cybersecurity indicators of compromise, other infrastructure, CVEs, or MITRE ATT\&CK Techniques. Querying this database based on the \emph{indicators} then allows analysts to find open source intelligence documents connected to that indicator and review their content; this graph database thus reduces the overhead required in searching a massive amount of open source documents to a succinct cloud of relevant documents. Importantly, this approach does not solve the intractable problem of unstructured open source text ingestion for the purposes of cybersecurity analysis. Our goal is to present one moderate-scale approach to this analysis task. Importantly, there are several limitations to the direct string matching of resources, namely because threat actors frequently change the infrastructure they use~\cite{metcalf2015blacklist, sood2013crimeware}. This then motivates more temporally oriented usage of the Neo4j graph database, but importantly in order for the system to be close to real time, the parsing and crawling infrastructure compute throughput would need to be improved substantially; this graph database was constructed continually over the course of months while the web crawlers were operating concurrently.

A good topic for future study is to have human analysts utilize the constructed Neo4j graph database on real world data in order to quantify the efficacy of the data, and the indicators that have been extracted from the documents. This is out of scope for the current study because of the time and resource intensive nature of such a study.

\section{Acknowledgements}
\label{section:acknowledgements}
Sandia National Laboratories is a multi-mission laboratory managed and operated by National Technology \& Engineering Solutions of Sandia, LLC (NTESS), a wholly owned subsidiary of Honeywell International Inc., for the U.S. Department of Energy’s National Nuclear Security Administration (DOE/NNSA) under contract DE-NA0003525. The New Mexico Cybersecurity Center of Excellence (NMCCoE) is a statewide Research and Public Service Project supported center for economic development, education, and research. The authors would like to thank both SNL and NMCCoE for funding and computing system access and support.


\setlength\bibitemsep{0pt}
\printbibliography

\appendix
\section{Portable Executable malware visualization}
\label{section:appendix_PE_visual}

\begin{figure*}[h!]
    \centering
    \includegraphics[width=0.32\textwidth]{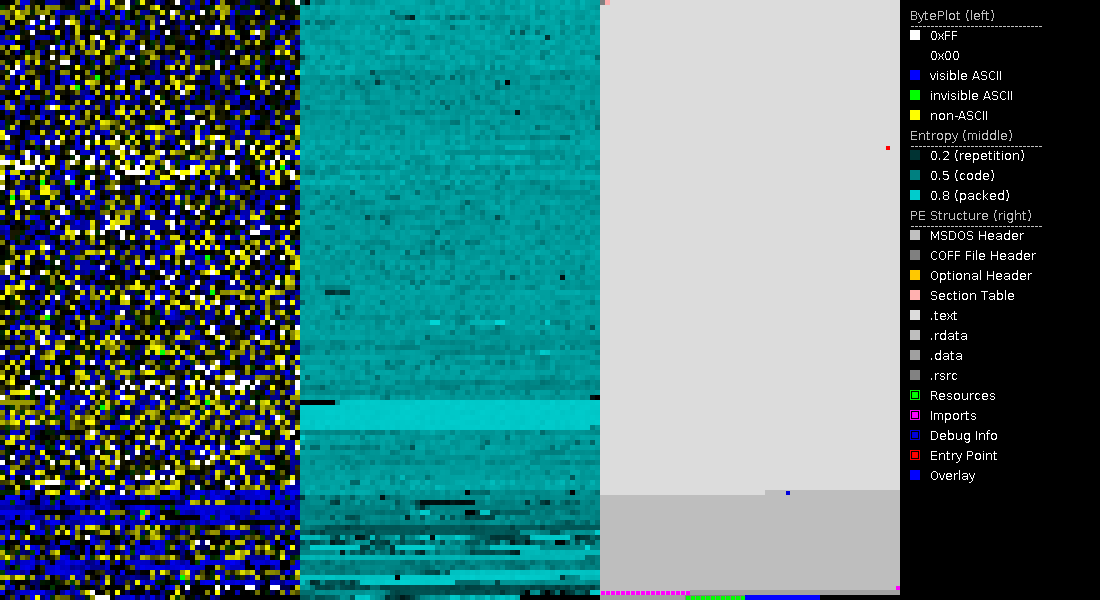}
    \includegraphics[width=0.32\textwidth]{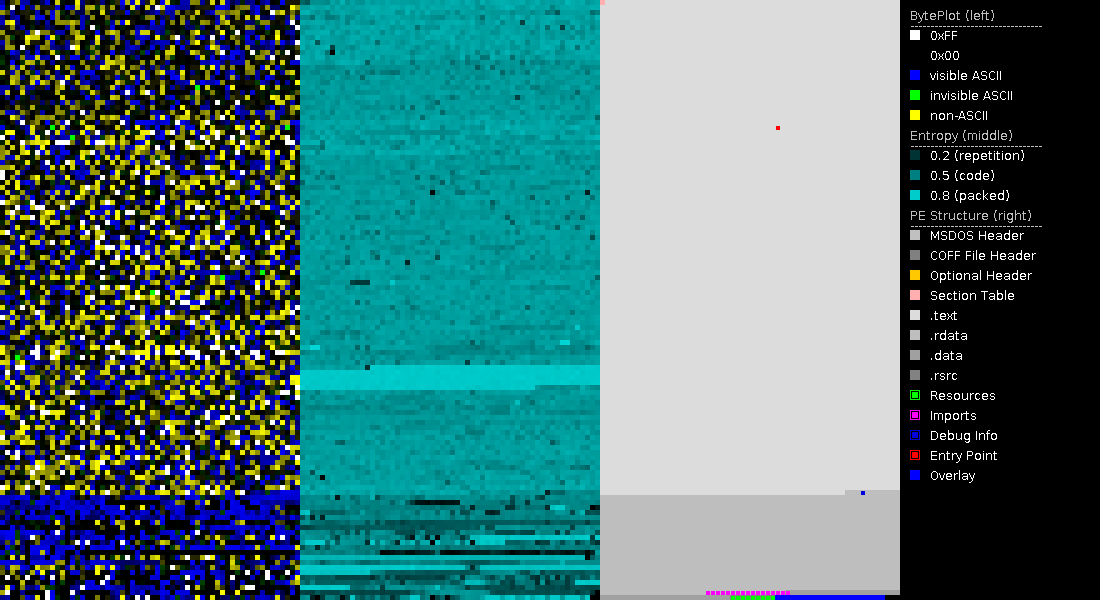}
    \includegraphics[width=0.32\textwidth]{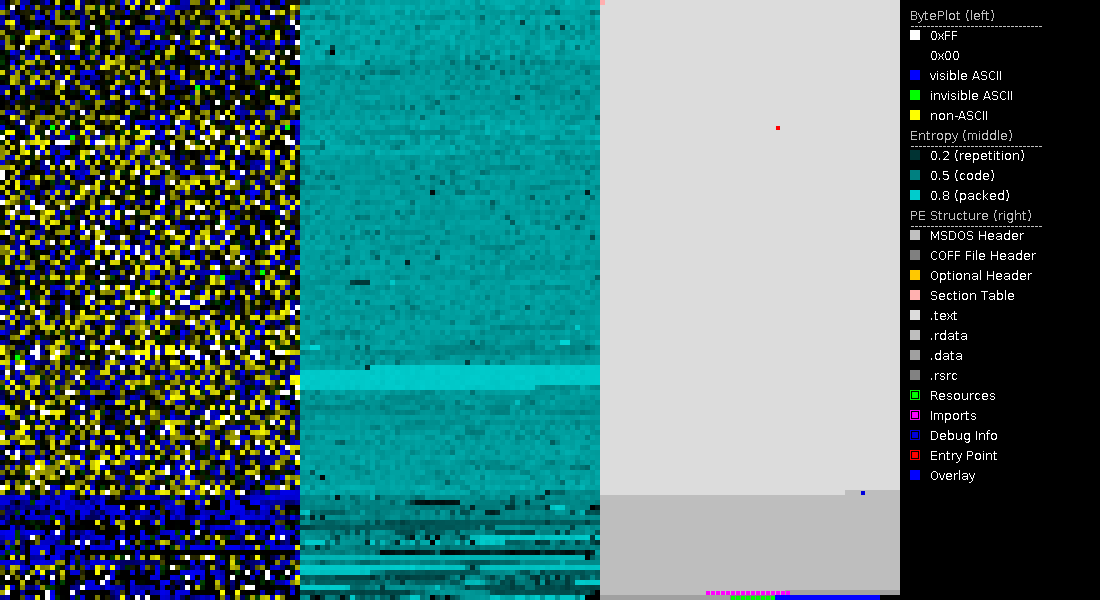}
    \caption{Visualization of three different PE32 malware samples byteplot, entropy, and structure. Each of these contain at least one common resource file (detailed in Section \ref{section:results_malware_PE_shared_resource}), which was sufficiently unique to link these pieces of malware. From left to right the sha256 checksums of these samples are \texttt{34aa24656d5527a5ff1f7eb4ce4e782085618ded3766730c81f8f16a15d7e0ce}, \texttt{4f0d5a81b8a5bc3f998a0ac7a37db5bad49e1a22173251ed916363375360d5a4}, \texttt{8b53201f1914764f384c6ec5a7a 5c5ab2924afaf382d2bbe79f68e43e5dfa3ba}. From left to right the \texttt{OriginalFilename} for each sample is \texttt{RKInstaller.exe}, \texttt{RKInstaller.exe}, and \texttt{POInstaller.exe}. Note the similar naming scheme to the four samples that are in Neo4j (see Section \ref{section:results_malware_PE_shared_resource}). 
    Interestingly, although these samples are not identical (we know this because their hashes are different), the middle and right hand side samples look visually indistinguishable from each other, whereas the left hand side sample is distinguishable from the other two. Overall, the structure of these PEs seem to be very similar which further indicates that these were likely developed by the same group or person. }
    \label{fig:PE_visuals}
\end{figure*}

Figure \ref{fig:PE_visuals} shows visualizations of three malware examples which contain a common PE resource artifact which was identified by its sha256 checksum. These three examples are different from the Neo4j nodes (specifically their hashes are not the same). These three distinct PE samples are also accessible on VirusTotal \footnote{\url{https://www.virustotal.com/gui/file/34aa24656d5527a5ff1f7eb4ce4e782085618ded3766730c81f8f16a15d7e0ce}} \footnote{\url{https://www.virustotal.com/gui/file/4f0d5a81b8a5bc3f998a0ac7a37db5bad49e1a22173251ed916363375360d5a4}} \footnote{\url{https://www.virustotal.com/gui/file/8b53201f1914764f384c6ec5a7a5c5ab2924afaf382d2bbe79f68e43e5dfa3ba}}. This suggests that this particular indicator, while unique and not very common, is likely seen on other static analysis tools beyond these two datasets. These visualizations were generated using FileScan\footnote{\url{https://www.filescan.io}} and PortEx\footnote{\url{https://github.com/struppigel/PortEx}}. 

\section{Large graph visualization}
\label{section:appendix_large_graphs}

\begin{figure*}[h!]
    \centering
    \includegraphics[width=0.49\textwidth]{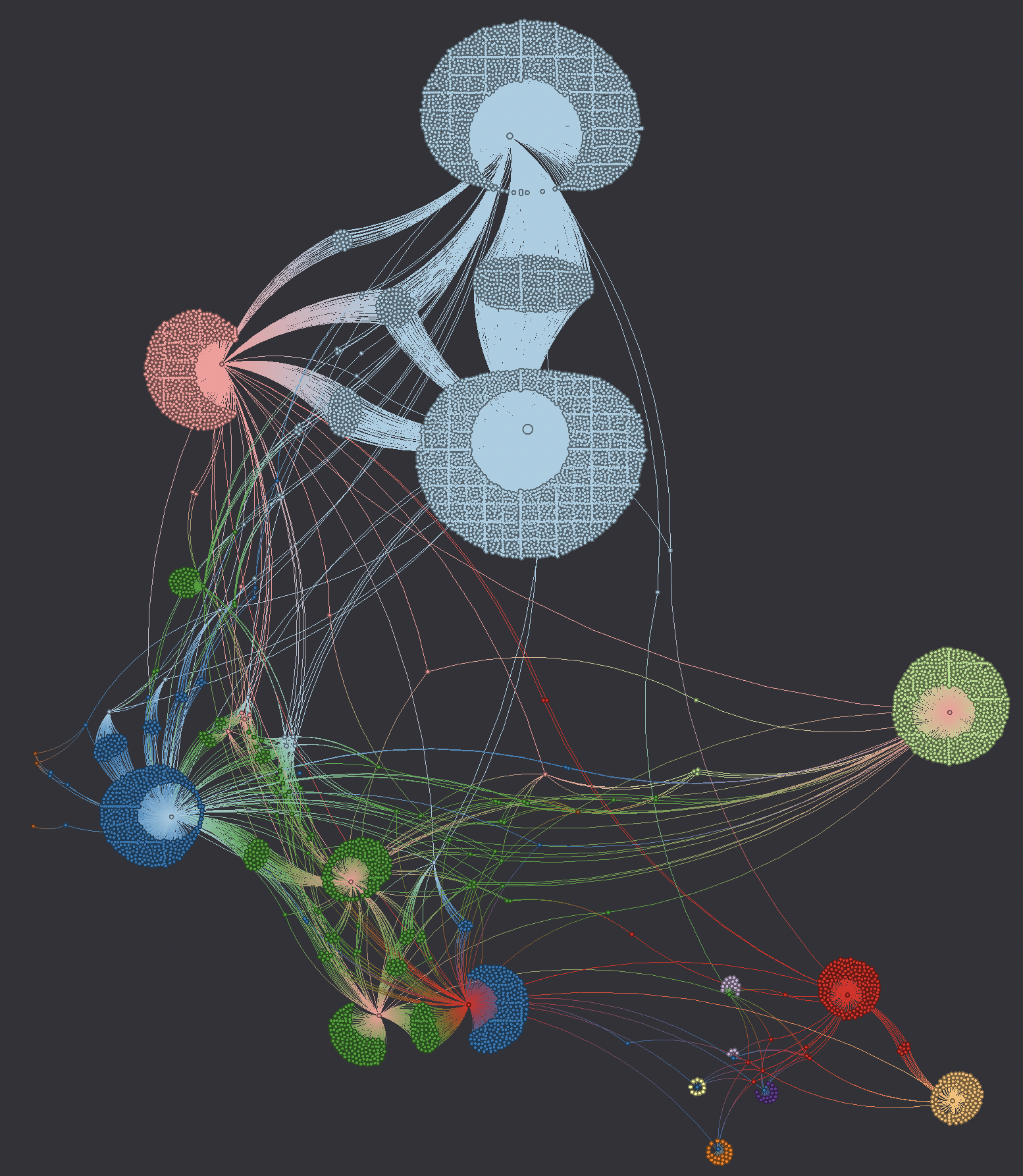}
    \includegraphics[width=0.49\textwidth]{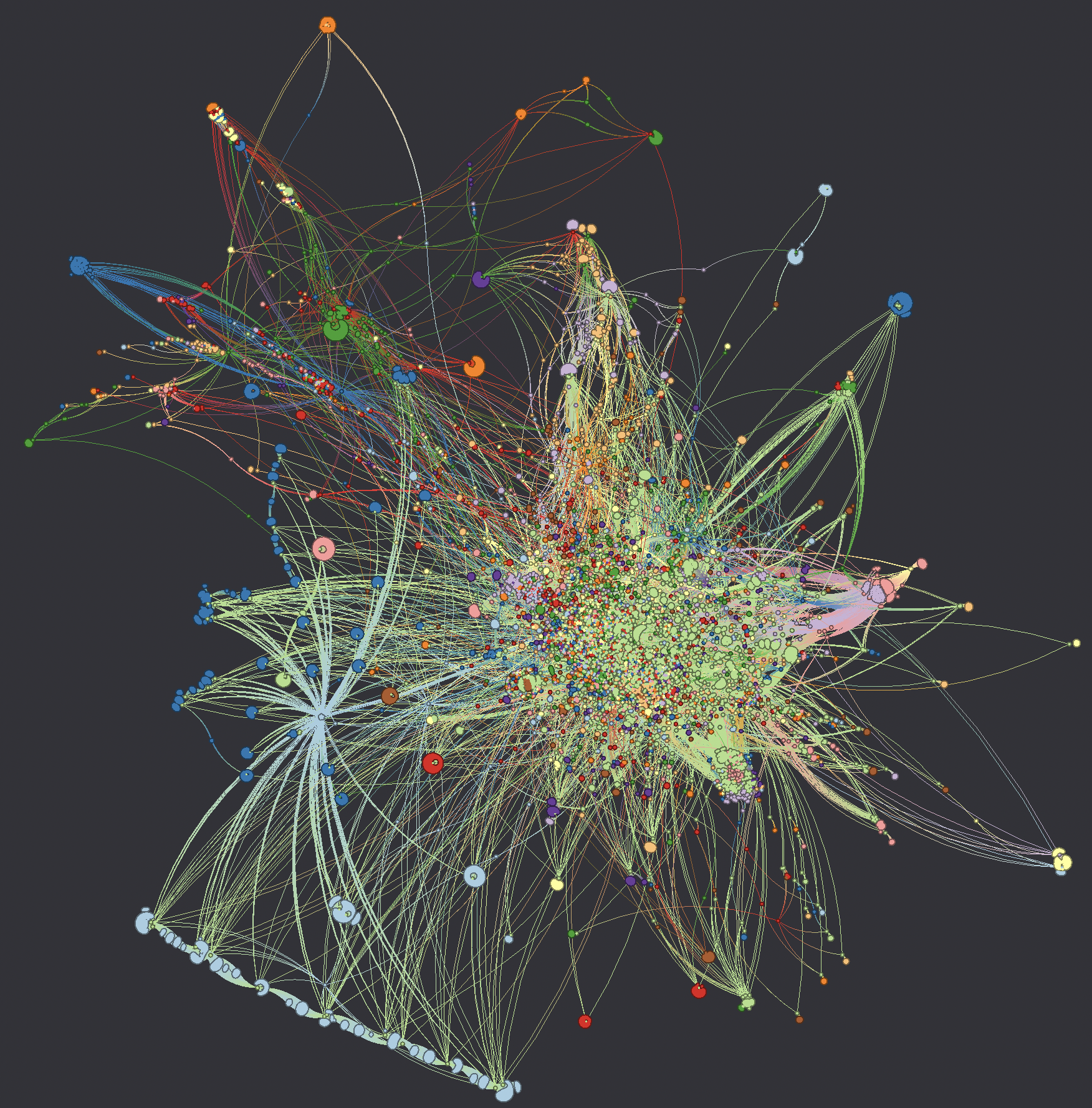}
    \caption{Large graph visualizations which are expansions of the Qakbot IP address graph from Figure \ref{fig:IP} in Section \ref{section:results_visual_IP_address}. Degree 3 connections (left) has 7,862 nodes and 9,244 edges, degree 4 connections (right) has 46,631 nodes and 112,935 edges. Note that the node and edge coloring is selected by the graphistry software and serves to delineate different aspects of the graph, but does not follow the same coloring scheme used in the Neo4j browser figures. These extremely large graph renderings show what the large scale behavior of this subgraph of the database looks like. Even though the database is overall quite sparse, there are clear clusters of the graph which behave similar to each other and there are also clearly very highly connected clusters. }
    \label{fig:large_graph_renderings}
\end{figure*}

Expanding out the network connections that exist in the Neo4j graph database is difficult to show visually because of the scale of the graphs in terms of edges and nodes. Here though in Figure \ref{fig:large_graph_renderings} we show two large graph examples, which is possible using pygraphistry\footnote{\url{https://github.com/graphistry/pygraphistry}} and networkx \cite{hagberg2008exploring} in python3. The graphistry version which was used to generate these figures is \texttt{2.39.32}, and the graph layout algorithm is \texttt{ForceAtlas2Barnes}. The primary observation that is important from Figure \ref{fig:large_graph_renderings} is that just a couple of degrees out from an indicator can already result in a very large graph, even though the entire graph database is very sparse. One of the reasons for this is that different types of indicators, names, and vulnerability IDs will naturally be mentioned at different rates in the gathered internet text. For example, popular malware names will be very common in cybersecurity text, whereas hashes will generally not be mentioned very frequently. Therefore, it is important to filter down to the types of edges you want to follow in the database when you are searching for specific indicators - for example by searching only for IP and domain edges if you are searching for server infrastructure connections. It is also useful in these cases to use the CTC metadata and language detection metadata in order to search specifically for document nodes which are, with high confidence, English text discussing cybersecurity in order to get more relevant text. When there is limited information available on an indicator or document, searching through all available network connections (such as in the malware hash cases shown in Sections \ref{section:results_visual_md5_hash} and \ref{section:results_malware_PE_shared_resource}) can also be useful. 

\end{document}